\documentclass[twocolumn,traditabstract,longauth]{aa}

\usepackage{graphicx,amsmath,amsfonts,amssymb}
\usepackage{epsf}
\usepackage{url}
\usepackage[breaklinks, colorlinks, citecolor=blue]{hyperref}
\usepackage{float}
\usepackage{hhline}
\usepackage{placeins}
\usepackage{array}
\newcolumntype{L}[1]{>{\raggedright\let\newline\\\arraybackslash\hspace{0pt}}m{#1}}
\usepackage{numprint}
\usepackage{tikz}
\usepackage{textpos}
\usepackage{natbib}

\bibpunct{(}{)}{;}{a}{}{,} 

\def\setsymbol#1#2{\expandafter\def\csname #1\endcsname{#2}}
\def\getsymbol#1{\csname #1\endcsname}

\def\Planck{\textit{Planck}}

\newbox\tablebox    \newdimen\tablewidth
\def\leaderfil{\leaders\hbox to 5pt{\hss.\hss}\hfil}
\def\endPlancktable{\tablewidth=\columnwidth 
    $$\hss\copy\tablebox\hss$$
    \vskip-\lastskip\vskip -2pt}
\def\endPlancktablewide{\tablewidth=\textwidth 
    $$\hss\copy\tablebox\hss$$
    \vskip-\lastskip\vskip -2pt}
\def\tablenote#1 #2\par{\begingroup \parindent=0.8em
    \abovedisplayshortskip=0pt\belowdisplayshortskip=0pt
    \noindent
    $$\hss\vbox{\hsize\tablewidth \hangindent=\parindent \hangafter=1 \noindent
    \hbox to \parindent{$^#1$\hss}\strut#2\strut\par}\hss$$
    \endgroup}
\def\doubleline{\vskip 3pt\hrule \vskip 1.5pt \hrule \vskip 5pt}

\def\L2{\ifmmode L_2\else $L_2$\fi}

\def\DeltaT{\ifmmode \Delta T\else $\Delta T$\fi}
\def\deltat{\ifmmode \Delta t\else $\Delta t$\fi}
\def\fknee{\ifmmode f_{\rm knee}\else $f_{\rm knee}$\fi}
\def\Fmax{\ifmmode F_{\rm max}\else $F_{\rm max}$\fi}
\def\solar{\ifmmode{\rm M}_{\mathord\odot}\else${\rm M}_{\mathord\odot}$\fi}
\def\Msolar{\ifmmode{\rm M}_{\mathord\odot}\else${\rm M}_{\mathord\odot}$\fi}
\def\Lsolar{\ifmmode{\rm L}_{\mathord\odot}\else${\rm L}_{\mathord\odot}$\fi}
\def\inv{\ifmmode^{-1}\else$^{-1}$\fi}
\def\mo{\ifmmode^{-1}\else$^{-1}$\fi}
\def\sup#1{\ifmmode ^{\rm #1}\else $^{\rm #1}$\fi}
\def\expo#1{\ifmmode \times 10^{#1}\else $\times 10^{#1}$\fi}
\def\,{\thinspace}
\def\lsim{\mathrel{\raise .4ex\hbox{\rlap{$<$}\lower 1.2ex\hbox{$\sim$}}}}
\def\gsim{\mathrel{\raise .4ex\hbox{\rlap{$>$}\lower 1.2ex\hbox{$\sim$}}}}

\def\simprop{\mathrel{\raise .4ex\hbox{\rlap{$\propto$}\lower 1.2ex\hbox{$\sim$}}}}
\def\deg{\ifmmode^\circ\else$^\circ$\fi}
\def\pdeg{\ifmmode $\setbox0=\hbox{$^{\circ}$}\rlap{\hskip.11\wd0 .}$^{\circ}
          \else \setbox0=\hbox{$^{\circ}$}\rlap{\hskip.11\wd0 .}$^{\circ}$\fi}
\def\arcs{\ifmmode {^{\scriptstyle\prime\prime}}
          \else $^{\scriptstyle\prime\prime}$\fi}
\def\arcm{\ifmmode {^{\scriptstyle\prime}}
          \else $^{\scriptstyle\prime}$\fi}
\newdimen\sa  \newdimen\sb
\def\parcs{\sa=.07em \sb=.03em
     \ifmmode \hbox{\rlap{.}}^{\scriptstyle\prime\kern -\sb\prime}\hbox{\kern -\sa}
     \else \rlap{.}$^{\scriptstyle\prime\kern -\sb\prime}$\kern -\sa\fi}
\def\parcm{\sa=.08em \sb=.03em
     \ifmmode \hbox{\rlap{.}\kern\sa}^{\scriptstyle\prime}\hbox{\kern-\sb}
     \else \rlap{.}\kern\sa$^{\scriptstyle\prime}$\kern-\sb\fi}
\def\ra[#1 #2 #3.#4]{#1\sup{h}#2\sup{m}#3\sup{s}\llap.#4}
\def\dec[#1 #2 #3.#4]{#1\deg#2\arcm#3\arcs\llap.#4}
\def\deco[#1 #2 #3]{#1\deg#2\arcm#3\arcs}
\def\rra[#1 #2]{#1\sup{h}#2\sup{m}}

\def\dots{\relax\ifmmode \ldots\else $\ldots$\fi}
\def\WHzsr{\ifmmode $W\,Hz\mo\,sr\mo$\else W\,Hz\mo\,sr\mo\fi}
\def\mHz{\ifmmode $\,mHz$\else \,mHz\fi}
\def\GHz{\ifmmode $\,GHz$\else \,GHz\fi}
\def\mKs{\ifmmode $\,mK\,s$^{1/2}\else \,mK\,s$^{1/2}$\fi}
\def\muKs{\ifmmode \,\mu$K\,s$^{1/2}\else \,$\mu$K\,s$^{1/2}$\fi}
\def\muKRJs{\ifmmode \,\mu$K$_{\rm RJ}$\,s$^{1/2}\else \,$\mu$K$_{\rm RJ}$\,s$^{1/2}$\fi}
\def\muKHz{\ifmmode \,\mu$K\,Hz$^{-1/2}\else \,$\mu$K\,Hz$^{-1/2}$\fi}
\def\MJysr{\ifmmode \,$MJy\,sr\mo$\else \,MJy\,sr\mo\fi}
\def\MJysrmK{\ifmmode \,$MJy\,sr\mo$\,mK$_{\rm CMB}\mo\else \,MJy\,sr\mo\,mK$_{\rm CMB}\mo$\fi}
\def\microns{\ifmmode \,\mu$m$\else \,$\mu$m\fi}

\def\muK{\ifmmode \,\mu$K$\else \,$\mu$\hbox{K}\fi}
\def\microK{\ifmmode \,\mu$K$\else \,$\mu$\hbox{K}\fi}
\def\muW{\ifmmode \,\mu$W$\else \,$\mu$\hbox{W}\fi}
\def\kms{\ifmmode $\,km\,s$^{-1}\else \,km\,s$^{-1}$\fi}
\def\kmsMpc{\ifmmode $\,\kms\,Mpc\mo$\else \,\kms\,Mpc\mo\fi}
\providecommand{\sorthelp}[1]{}

\renewcommand{\aap}{A\&A}

\renewcommand{\apjl}{ApJL}

\newcommand{\pic}{\planck\ input catalogue}

\newcommand{\planck}{\textit{Planck}}
\newcommand{\pmnt}{{PCNT}}
\newcommand{\bpmnt}{{PCNTb}}
\newcommand{\hpmnt}{{PCNThs}}
\def\GAL#1{{GAL#1}}

\providecommand{\sorthelp}[1]{}

\begin{document}
\author{\small
Planck Collaboration: Y.~Akrami\inst{51, 53}
\and
F.~Arg\"{u}eso\inst{17}
\and
M.~Ashdown\inst{60, 6}
\and
J.~Aumont\inst{84}
\and
C.~Baccigalupi\inst{70}
\and
M.~Ballardini\inst{21, 37}
\and
A.~J.~Banday\inst{84, 9}
\and
R.~B.~Barreiro\inst{55}
\and
N.~Bartolo\inst{26, 56}
\and
S.~Basak\inst{76}
% ...
}
\institute{\small
\goodbreak
\and
AIM, CEA, CNRS, Universit\'{e} Paris-Saclay, Universit\'{e} Paris-Diderot, Sorbonne Paris Cit\'{e}, F-91191 Gif-sur-Yvette, France\goodbreak
\and
APC, AstroParticule et Cosmologie, Universit\'{e} Paris Diderot, CNRS/IN2P3, CEA/lrfu, Observatoire de Paris, Sorbonne Paris Cit\'{e}, 10, rue Alice Domon et L\'{e}onie Duquet, 75205 Paris Cedex 13, France\goodbreak
\and
Aalto University Mets\"ahovi Radio Observatory and Dept of Electronics and Nanoengineering, P.O. Box 15500, FI-00076 AALTO, Finland\goodbreak
\and
African Institute for Mathematical Sciences, 6-8 Melrose Road, Muizenberg, Cape Town, South Africa\goodbreak
\and
Astrophysics Group, Cavendish Laboratory, University of Cambridge, J J Thomson Avenue, Cambridge CB3 0HE, U.K.\goodbreak
\and
Astrophysics \& Cosmology Research Unit, School of Mathematics, Statistics \& Computer Science, University of KwaZulu-Natal, Westville Campus, Private Bag X54001, Durban 4000, South Africa\goodbreak
\and
CITA, University of Toronto, 60 St. George St., Toronto, ON M5S 3H8, Canada\goodbreak
\and
CNRS, IRAP, 9 Av. colonel Roche, BP 44346, F-31028 Toulouse cedex 4, France\goodbreak
\and
Cahill Center for Astronomy and Astrophysics, California Institute of Technology, Pasadena CA,  91125, USA\goodbreak
\and
California Institute of Technology, Pasadena, California, U.S.A.\goodbreak
\and
Computational Cosmology Center, Lawrence Berkeley National Laboratory, Berkeley, California, U.S.A.\goodbreak
\and
DTU Space, National Space Institute, Technical University of Denmark, Elektrovej 327, DK-2800 Kgs. Lyngby, Denmark\goodbreak
\and
D\'{e}partement de Physique Th\'{e}orique, Universit\'{e} de Gen\`{e}ve, 24, Quai E. Ansermet,1211 Gen\`{e}ve 4, Switzerland\goodbreak
\and
Departamento de Astrof\'{i}sica, Universidad de La Laguna (ULL), E-38206 La Laguna, Tenerife, Spain\goodbreak
\and
Departamento de F\'{\i}sica, Universidad de Oviedo, C/ Federico Garc\'{\i}a Lorca, 18 , Oviedo, Spain\goodbreak
\and
Departamento de Matem\'{a}ticas, Universidad de Oviedo, C/ Federico Garc\'{\i}a Lorca, 18, Oviedo, Spain\goodbreak
\and
Department of Astrophysics/IMAPP, Radboud University, P.O. Box 9010, 6500 GL Nijmegen, The Netherlands\goodbreak
\and
Department of Mathematics, University of Stellenbosch, Stellenbosch 7602, South Africa\goodbreak
\and
Department of Physics \& Astronomy, University of British Columbia, 6224 Agricultural Road, Vancouver, British Columbia, Canada\goodbreak
\and
Department of Physics \& Astronomy, University of the Western Cape, Cape Town 7535, South Africa\goodbreak
\and
Department of Physics and Astronomy, University of Sussex, Brighton BN1 9QH, U.K.\goodbreak
\and
Department of Physics, Gustaf H\"{a}llstr\"{o}min katu 2a, University of Helsinki, Helsinki, Finland\goodbreak
\and
Department of Physics, Princeton University, Princeton, New Jersey, U.S.A.\goodbreak
\and
Department of Physics, University of California, Santa Barbara, California, U.S.A.\goodbreak
\and
Dipartimento di Fisica e Astronomia G. Galilei, Universit\`{a} degli Studi di Padova, via Marzolo 8, 35131 Padova, Italy\goodbreak
\and
Dipartimento di Fisica e Scienze della Terra, Universit\`{a} di Ferrara, Via Saragat 1, 44122 Ferrara, Italy\goodbreak
\and
Dipartimento di Fisica, Universit\`{a} La Sapienza, P. le A. Moro 2, Roma, Italy\goodbreak
\and
Dipartimento di Fisica, Universit\`{a} degli Studi di Milano, Via Celoria, 16, Milano, Italy\goodbreak
\and
Dipartimento di Fisica, Universit\`{a} degli Studi di Trieste, via A. Valerio 2, Trieste, Italy\goodbreak
\and
Dipartimento di Fisica, Universit\`{a} di Roma Tor Vergata, Via della Ricerca Scientifica, 1, Roma, Italy\goodbreak
\and
European Space Agency, ESAC, Planck Science Office, Camino bajo del Castillo, s/n, Urbanizaci\'{o}n Villafranca del Castillo, Villanueva de la Ca\~{n}ada, Madrid, Spain\goodbreak
\and
European Space Agency, ESTEC, Keplerlaan 1, 2201 AZ Noordwijk, The Netherlands\goodbreak
\and
Gran Sasso Science Institute, INFN, viale F. Crispi 7, 67100 L'Aquila, Italy\goodbreak
\and
Haverford College Astronomy Department, 370 Lancaster Avenue, Haverford, Pennsylvania, U.S.A.\goodbreak
\and
Helsinki Institute of Physics, Gustaf H\"{a}llstr\"{o}min katu 2, University of Helsinki, Helsinki, Finland\goodbreak
\and
INAF - OAS Bologna, Istituto Nazionale di Astrofisica - Osservatorio di Astrofisica e Scienza dello Spazio di Bologna, Area della Ricerca del CNR, Via Gobetti 101, 40129, Bologna, Italy\goodbreak
\and
INAF - Osservatorio Astronomico di Padova, Vicolo dell'Osservatorio 5, Padova, Italy\goodbreak
\and
INAF - Osservatorio Astronomico di Trieste, Via G.B. Tiepolo 11, Trieste, Italy\goodbreak
\and
INAF, Istituto di Radioastronomia, Via Piero Gobetti 101, I-40129 Bologna, Italy\goodbreak
\and
INAF/IASF Milano, Via E. Bassini 15, Milano, Italy\goodbreak
\and
INFN, Sezione di Bologna, viale Berti Pichat 6/2, 40127 Bologna, Italy\goodbreak
\and
INFN, Sezione di Ferrara, Via Saragat 1, 44122 Ferrara, Italy\goodbreak
\and
INFN, Sezione di Milano, Via Celoria 16, Milano, Italy\goodbreak
\and
INFN, Sezione di Roma 1, Universit\`{a} di Roma Sapienza, Piazzale Aldo Moro 2, 00185, Roma, Italy\goodbreak
\and
INFN, Sezione di Roma 2, Universit\`{a} di Roma Tor Vergata, Via della Ricerca Scientifica, 1, Roma, Italy\goodbreak
\and
Imperial College London, Astrophysics group, Blackett Laboratory, Prince Consort Road, London, SW7 2AZ, U.K.\goodbreak
\and
Infrared Processing and Analysis Center, California Institute of Technology, Pasadena, CA 91125, U.S.A.\goodbreak
\and
Institut d'Astrophysique Spatiale, CNRS, Univ. Paris-Sud, Universit\'{e} Paris-Saclay, B\^{a}t. 121, 91405 Orsay cedex, France\goodbreak
\and
Institut d'Astrophysique de Paris, CNRS (UMR7095), 98 bis Boulevard Arago, F-75014, Paris, France\goodbreak
\and
Institute Lorentz, Leiden University, PO Box 9506, Leiden 2300 RA, The Netherlands\goodbreak
\and
Institute of Astronomy, University of Cambridge, Madingley Road, Cambridge CB3 0HA, U.K.\goodbreak
\and
Institute of Theoretical Astrophysics, University of Oslo, Blindern, Oslo, Norway\goodbreak
\and
Instituto de Astrof\'{\i}sica de Canarias, C/V\'{\i}a L\'{a}ctea s/n, La Laguna, Tenerife, Spain\goodbreak
\and
Instituto de F\'{\i}sica de Cantabria (CSIC-Universidad de Cantabria), Avda. de los Castros s/n, Santander, Spain\goodbreak
\and
Istituto Nazionale di Fisica Nucleare, Sezione di Padova, via Marzolo 8, I-35131 Padova, Italy\goodbreak
\and
Jet Propulsion Laboratory, California Institute of Technology, 4800 Oak Grove Drive, Pasadena, California, U.S.A.\goodbreak
\and
Jodrell Bank Centre for Astrophysics, Alan Turing Building, School of Physics and Astronomy, The University of Manchester, Oxford Road, Manchester, M13 9PL, U.K.\goodbreak
\and
Kavli Institute for Cosmological Physics, University of Chicago, Chicago, IL 60637, USA\goodbreak
\and
Kavli Institute for Cosmology Cambridge, Madingley Road, Cambridge, CB3 0HA, U.K.\goodbreak
\and
Kavli Institute for the Physics and Mathematics of the Universe (Kavli IPMU, WPI), UTIAS, The University of Tokyo, Chiba, 277- 8583, Japan\goodbreak
\and
Laboratoire de Physique Subatomique et Cosmologie, Universit\'{e} Grenoble-Alpes, CNRS/IN2P3, 53, rue des Martyrs, 38026 Grenoble Cedex, France\goodbreak
\and
Laboratoire de Physique Th\'{e}orique, Universit\'{e} Paris-Sud 11 \& CNRS, B\^{a}timent 210, 91405 Orsay, France\goodbreak
\and
Lawrence Berkeley National Laboratory, Berkeley, California, U.S.A.\goodbreak
\and
Low Temperature Laboratory, Department of Applied Physics, Aalto University, Espoo, FI-00076 AALTO, Finland\goodbreak
\and
Max-Planck-Institut f\"{u}r Astrophysik, Karl-Schwarzschild-Str. 1, 85741 Garching, Germany\goodbreak
\and
Mullard Space Science Laboratory, University College London, Surrey RH5 6NT, U.K.\goodbreak
\and
NAOC-UKZN Computational Astrophysics Centre (NUCAC), University of KwaZulu-Natal, Durban 4000, South Africa\goodbreak
\and
Nicolaus Copernicus Astronomical Center, Polish Academy of Sciences, Bartycka 18, 00-716 Warsaw, Poland\goodbreak
\and
SISSA, Astrophysics Sector, via Bonomea 265, 34136, Trieste, Italy\goodbreak
\and
San Diego Supercomputer Center, University of California, San Diego, 9500 Gilman Drive, La Jolla, CA 92093, USA\goodbreak
\and
School of Chemistry and Physics, University of KwaZulu-Natal, Westville Campus, Private Bag X54001, Durban, 4000, South Africa\goodbreak
\and
School of Physical Sciences, National Institute of Science Education and Research, HBNI, Jatni-752050, Odissa, India\goodbreak
\and
School of Physics and Astronomy, Cardiff University, Queens Buildings, The Parade, Cardiff, CF24 3AA, U.K.\goodbreak
\and
School of Physics and Astronomy, Sun Yat-sen University, 2 Daxue Rd, Tangjia, Zhuhai, China\goodbreak
\and
School of Physics, Indian Institute of Science Education and Research Thiruvananthapuram, Maruthamala PO, Vithura, Thiruvananthapuram 695551, Kerala, India\goodbreak
\and
Simon Fraser University, Department of Physics, 8888 University Drive, Burnaby BC, Canada\goodbreak
\and
Sorbonne Universit\'{e}, Observatoire de Paris, Universit\'{e} PSL, \'{E}cole normale sup\'{e}rieure, CNRS, LERMA, F-75005, Paris, France\goodbreak
\and
Sorbonne Universit\'{e}-UPMC, UMR7095, Institut d'Astrophysique de Paris, 98 bis Boulevard Arago, F-75014, Paris, France\goodbreak
\and
Space Science Data Center - Agenzia Spaziale Italiana, Via del Politecnico snc, 00133, Roma, Italy\goodbreak
\and
Space Sciences Laboratory, University of California, Berkeley, California, U.S.A.\goodbreak
\and
The Oskar Klein Centre for Cosmoparticle Physics, Department of Physics, Stockholm University, AlbaNova, SE-106 91 Stockholm, Sweden\goodbreak
\and
UPMC Univ Paris 06, UMR7095, 98 bis Boulevard Arago, F-75014, Paris, France\goodbreak
\and
Universit\'{e} de Toulouse, UPS-OMP, IRAP, F-31028 Toulouse cedex 4, France\goodbreak
% ...
}

\title{\textit{Planck} intermediate results. LIV.\\
The Planck Multi-frequency Catalogue of Non-thermal Sources}

\titlerunning{Planck Multi-frequency Catalogue of Non-thermal Sources}


\abstract{This paper presents the Planck Multi-frequency Catalogue of
Non-thermal (i.e. synchrotron-dominated) Sources (\pmnt)
observed between 30 and 857\,GHz by the ESA \planck\ mission.
This catalogue was constructed by selecting objects detected in the
full mission all-sky temperature maps at 30 and 143\,GHz, with a
signal-to-noise ratio (S/N) $> 3$ in at least one of the two channels after
filtering with a particular Mexican hat wavelet. As a result, 29\,400
source candidates were selected. Then, a multi-frequency analysis
was performed using the Matrix Filters methodology at the position
of these objects, and flux densities and errors were calculated for all
of them in the nine \planck\ channels. This catalogue was built using a
different methodology than the one adopted for the Planck Catalogue of
Compact Sources (PCCS) and the Second Planck Catalogue of Compact Sources
(PCCS2), although the initial detection was done with the same pipeline that
was used to produce them. The present catalogue is the first unbiased,
full-sky catalogue of synchrotron-dominated sources published at millimetre
and submillimetre wavelengths and constitutes a powerful database for
statistical studies of non-thermal extragalactic sources, whose emission
is dominated by the central active galactic nucleus.
Together with the full multi-frequency catalogue, we also define the
 Bright Planck Multi-frequency Catalogue of Non-thermal Sources
({\bpmnt}), where only those objects with a ${\rm S/N}>4$ at both 30 and
143\,GHz were selected. In this catalogue 1146 compact
sources are detected outside the adopted \planck\ GAL070 mask;
thus, these sources constitute a highly reliable sample of extragalactic
radio sources. We also flag the  high-significance subsample
({\hpmnt}), a subset of 151 sources that are detected with ${\rm S/N}>4$
in all nine \planck\ channels, 75 of which are found outside the \planck\
mask adopted here.  The remaining 76 sources inside the Galactic mask are
very likely Galactic objects.}


\maketitle

%

\section{Introduction} \label{sec:intro}

This paper, one of a series associated with the 2015 release of data from the
\planck\footnote{\planck\ (\url{http://www.esa.int/Planck}) is a
project of the European Space Agency (ESA) with instruments provided by two scientific consortia funded by ESA
member states and led by Principal Investigators from France and Italy, telescope reflectors provided through a
collaboration between ESA and a scientific consortium led and funded by Denmark, and additional contributions from
NASA (USA).}
 mission \citep{planck2014-a01},
outlines the construction of the first full-sky multi-frequency catalogue of non-thermal, i.e. synchrotron-dominated sources detected in \planck\ full-mission \citep{planck2014-a01} temperature maps. The main
purpose of this catalogue is to provide a full-sky sample of extragalactic radio sources (ERSs), of which the majority
 show a flat ($\alpha\,{\simeq}\,0$) observed spectral emission distribution\footnote{Here we follow the convention $S(\nu)\propto \nu^{\alpha}$, with $S(\nu)$
the observed flux density at frequency $\nu$ and $\alpha$ the spectral index.} up to sub-mm/far-IR wavelengths.
It thus extends to higher frequencies than had been achieved with large-area ground-based surveys (NVSS, FIRST, GB6, Parkes,
AT20G, etc.), which are currently limited to centimetre wavelengths. Unlike  previous \planck\ catalogues that were
constructed on a frequency-by-frequency basis, this is a fully multi-frequency catalogue of point sources obtained
through a multi-band filtering technique, which uses data from both the
Low Frequency Instrument \cite[LFI,][]{planck2014-a07} and
High Frequency Instrument \citep[HFI,][]{planck2014-a09}.
In addition to the aforementioned ERSs, this catalogue also includes
several thousands of candidate compact sources of Galactic origin.

As extensively discussed in previous papers that studied the ERS populations observed at millimetre wavelengths
\citep[e.g.][]{zotti05,zotti10,tucci11,planck2011-6.1,planck2011-6.2,planck2011-6.3a,planck2012-VII,zotti15,planck2016-XLV} the majority of these extragalactic sources are
usually classified as flat-spectrum radio quasars (FSRQ) and BL Lac objects, collectively called
blazars,
and only a few of them are classified as inverted spectrum or
high-frequency peaker (HFP) radio sources.\footnote{It is widely agreed that HFP sources correspond to the early
stages of evolution of powerful radio sources, when the radio emitting region grows and expands in the interstellar
medium of the host galaxy before becoming an extended radio source \citep[see  e.g.][]{odea98,zotti05}.}

Blazars are a relatively rare class of active galactic nuclei (AGN) characterized by electromagnetic emission
over the entire energy spectrum, from the radio band to the most energetic gamma rays \citep[e.g.][]{planck2011-6.3b}. They are also
characterized by highly variable, non-thermal synchrotron emission with relatively high linear polarization at
GHz frequencies \citep{moore81} in which the beamed component dominates the observed emission
\citep{angel80}. Interestingly, observations of blazars at mm/sub-mm wavelengths often reveal the transition from
optically thick to optically thin radio emission in the most compact regions, i.e. they provide information on the maximum
self-absorption frequency and thus of the physical dimension, $r_{\rm M}$, of the self-absorbed (optically thick) core
region
\citep{konigl81,tucci11}. Their spectral energy distribution (SED) is characterized by two
broad peaks in $\nu L_{\nu}$: the first peak, attributed to Doppler-boosted synchrotron radiation, occurs at a
frequency $\nu_{\rm p}$ varying from $10^{12}$ to $10^{18}\,$Hz
\citep{niep06}; and
the second peak, attributed to inverse Compton scattering, occurs at high gamma ray energies, up to around $10^{25}\,$Hz
\citep[see e.g.][]{planck2011-6.3b}.
Because of this second emission peak, blazars constitute a numerous class of extragalactic gamma ray sources.
About $90\,\%$ of the firmly
identified extragalactic sources, and about $94\,\%$ of the associated
sources (i.e. of sources having a counterpart with a $> 80\,\%$
probability of being the real identification) in the Third Fermi
Large Area Telescope (LAT) catalogue, are blazars \citep{3LAC}.
 Thus, knowledge of the blazar population
is also very relevant in high-energy astrophysics. Furthermore, because of their very broad spectral emission,
blazars are expected to contribute a substantial
fraction of the extragalactic background
 and its fluctuations at both millimetre wavelengths and
 very high energies, i.e. gamma rays.
For all these reasons, this new full-sky multi-frequency catalogue of non-thermal synchrotron ERSs -- aimed
at filling, albeit partially, the gap between samples selected
at mid/near-IR wavelengths 
\citep[e.g.][]{WISEp}
and those collected from space observations at
very short wavelengths ({\it Compton}-EGRET, {\it Fermi}-LAT, Swift/BAT, etc.) -- will not only allow the identification of
new blazars to be included in multi-frequency catalogues
\citep[e.g.][]{massaro09,massaro15},
 but
 will also help future studies of the blazar phenomenon, of the physical processes occurring in the nuclear region of
this class of sources, and of their cosmological evolution.

The outline of the paper is as follows. In Sect.~\ref{sec:cat0} we describe the criteria we follow to select
radio source candidates at 30 and 143\,GHz by using a blind detection on \planck\ maps filtered with the Mexican hat wavelet. In Sect.~\ref{sec:multifreq} we briefly review the multi-frequency detection method we use to construct our
catalogue and we give details about the implementation of this technique on the \planck\ data. The mathematical description of the method is 
referred to the Appendix~\ref{sec:appendix}, while the practicalities of the implementation of the multi-frequency detection technique for \planck\, data are
discussed in Appendix~\ref{sec:practical}.
 The catalogue is
introduced in Sect.~\ref{sec:cat1}. We also define and introduce in Sect.~\ref{sec:cat1} the Bright Planck Multi-frequency Catalogue of
Non-thermal Sources ({\bpmnt}) and also the high-significance subsample ({\hpmnt}). 
The internal \planck\ and the external validation of the
catalogues and a brief description of their statistical properties are discussed in Sect.~\ref{sec:valprop}. 
The released catalogue is described in Sect.~\ref{sec:contents}.
Finally, we summarize our main
conclusions in Sect.~\ref{sec:conclusions}.

\section{The input \planck\ sample} \label{sec:cat0}

Our list of source candidates is the union of those detected with ${\rm S/N}> 3$ in either the 30- and 143-GHz
 maps in total intensity made by \planck.  The maps used here are the full-mission ones and
cover the entire sky \citep{planck2014-a01}. We employ two frequencies to detect sources because our
multi-frequency analysis typically yields lower flux density detection limits for the same S/N than
a single frequency approach (see Sect.~\ref{sec:multifreq}). Thus the completeness is improved, without
compromising reliability.  This multi-frequency approach allows us to reduce the single-frequency threshold of ${\rm S/N}> 4$ used for
the PCCS2 \citep{planck2014-a35}, thus allowing additional candidates to enter the catalogue.
Our choice of two \planck\ bands at 30 and 143\,GHz was made to combine the power of observations at low radio
frequency (where synchrotron emission is typically dominant in ERS spectra), with the low-noise and
minimum-foreground 143-GHz \planck\ band. Note that ERSs still dominate the source counts at 143\,GHz, as displayed
and discussed, e.g. in figure~25 of the PCCS2 paper \citep{planck2014-a35} and also in figure~10 of
\citep{planck2012-VII}.  Had we only selected sources at 30\,GHz, we would have taken the risk of losing
inverted spectrum sources, whose spectra can be rising well above 30\,GHz, as well as high-frequency peakers \citep[HFPs,
see for example ][]{HFP}, which display a very distinct spectral emission, peaking up to millimetre wavelengths.
Given the main purpose of this paper -- i.e. the selection of a full-sky sample of non-thermal ERSs -- the adopted
selection criteria help to ensure that we are not defining a sample/catalogue that is {\it biased} in origin against some
particular ERS population, detectable in principle by the \planck\ full-sky surveys.

The pipeline used to produce the two initial lists of source candidates at 30 and 143\,GHz with ${\rm S/N}>3$ is the
same one used to produce the PCCS \citep{planck2013-p05} and PCCS2 \citep{planck2014-a35}
low frequency catalogues. For convenience, we use \planck\ maps upgraded to a common resolution 
parameter $N_{\rm side} = 2048$ \citep[using \texttt{HEALPix}, ][]{gorski2005}.
One of the characteristics of this pipeline
is a two-step analysis, first performing a blind detection in the full-sky map, and, second, repeating the analysis
in a non-blind fashion at the positions of the sources detected in the first step. The goal of this multi-step
analysis is to reduce the number of spurious detections introduced by the filtering approach in the borders of the cut-out
patches, and to measure the flux density of sources and background noise in an optimal way. In practice, in the first step
the sky is projected onto a sufficient number of overlapping square flat patches ($7\pdeg33 \times 7\pdeg33$, $128
\times 128$ pixels in size) such that even after removing the sources detected near the edges of the patch, the full
sky was effectively analysed. Second, at the position of the remaining source candidates, a new patch (with
the same dimensions) is constructed, redoing the filtering and optimization of the Mexican hat wavelet scale, but this
time focusing on the centre of the image, re-assessing the S/N of the source. Sources that in this second step
do not meet the ${\rm S/N}>3$ criterion are rejected. Since the patches overlap in order to smoothly cover the entire sky \citep[see the details in][]{planck2013-p05,planck2014-a35}, multiple entries of the same
source can be obtained in overlapping regions of two or more sky patches.
These multiple entries are removed from the catalogue, keeping only those with the highest S/N in each case.

The \planck\ 143-GHz channel has much better angular resolution than the 30-GHz channel. Therefore, it can happen that two or more objects in our \planck\ sample, selected at 143\,GHz, are inside the same 30\,GHz beam solid angle. We keep such multiple occurrences in our input sample and deal with them in a way that will be described in Sect.~\ref{sec:description} and the Appendix~\ref{sec:blend}.

Finally, the main disadvantage of selecting sources at 143\,GHz is the possibility of unintentionally selecting Galactic (or extragalactic) cold thermal sources that are bright enough to pass our selection threshold. 
We will try to quantify the impact of these sources in Sect.~\ref{sec:Galactic}. 

\section{Multi-frequency detection} \label{sec:multifreq}

The vast majority of component-separation methods that are typically used in CMB experiments take advantage of the \emph{spectral diversity} of the different astrophysical components (synchrotron, thermal dust, the CMB itself, etc.) as a means to disentangle the various  signals that arrive at the detectors \citep[see for example][]{leach2008,planck2014-a11,planck2014-a12}. However, most of the  techniques
that are used for the separation of diffuse components are not well suited for  the detection of extragalactic compact sources, except  for the particular case of galaxy clusters observed through the thermal Sunyaev-Zeldovich (tSZ) effect. Individual galaxies   
leave their imprint on the microwave sky through an enormous variety of astrophysical mechanisms -- from radio active lobes to dust thermal emission -- so that, strictly speaking, each individual galaxy has its own unique spectral behaviour, thus making it impossible for methods that rely on spectral diversity alone to 
solve for each individual spectral signature.
New methods, specifically tailored for compact sources, then become necessary. Most of the detection methods for compact-source detection that have been proposed in the literature  make use of the \emph{scale diversity} instead of spectral diversity (that is, they focus on the sizes of the sources rather than on their colours). In most cases, the catalogues of  extragalactic sources are extracted from CMB maps separately, one frequency channel at a time. The previously published \planck\ catalogues \citep{planck2011-1.10,planck2013-p05,planck2014-a35} follow this single-frequency approach.

Methods that attempt to combine the spectral and scale-diversity principles in order to 
enhance the detectability of extragalactic compact sources were fist introduced in the literature in the context of the study of the tSZ effect \citep{herr02,herr05,melin06,planck2011-5.1a,planck2013-p05a,planck2013-p05a-addendum,planck2014-a36} an were afterwards generalized to generic compact-source populations \citep{lanz10,lanz13}.
The matched matrix filters (MTXFs) approach was introduced in \cite{MTXFa} and further explored in \cite{MTXFb} as a solution to the problem of multi-frequency detection of extragalactic point sources when the frequency dependence of the sources is not known a priori. 
The MTXFs have already been used inside the Planck Collaboration to validate the LFI part of the PCCS catalogue \citep{planck2013-p05} and we will use them in this paper.  The main aspects of MTXF theory can be found in \cite{MTXFa} and \citet{MTXFb}, and are also reviewed in the general context of point source detection in CMB experiments in \cite{herranz10} and \cite{multireview}; however, for clarity, we summarize the main
mathematical aspects of MTXF theory in Appendix~\ref{sec:appendix}. The practical details of the implementation of the MTXFs for \planck\, data are described in Appendix~\ref{sec:practical}. In particular, we discuss how we obtain the photometry of the sources in section~\ref{sec:photometry} of Appendix~\ref{sec:practical}.

\section{The Planck Multi-frequency Catalogue of Non-thermal Sources} \label{sec:cat1}

\subsection{General description} \label{sec:description}

\begin{figure*}[h]
  \centering
   \includegraphics[width=\textwidth]{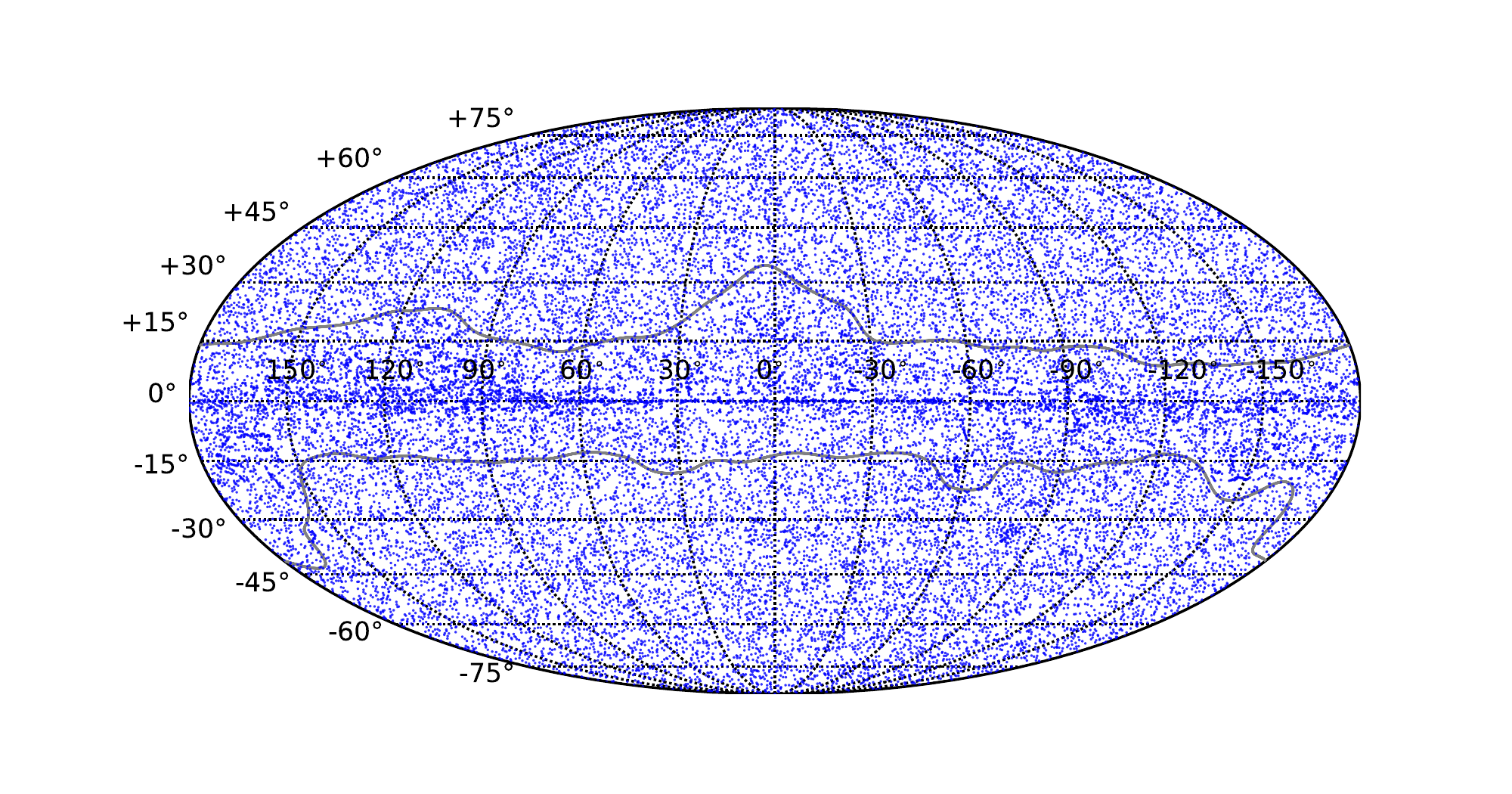}
   \caption{Position on the sky (Galactic coordinates) of the {\pmnt} sources. The boundary of the \planck\ $70\,\%$ Galactic mask, \GAL070, is superimposed in grey. }
              \label{fig:positions_PMNT}
\end{figure*}

We provide a Planck Multi-frequency Catalogue of Non-thermal Sources ({\pmnt})
 containing the 29\,400 candidates selected in our input catalogue (as described in Sect.~\ref{sec:cat0}); additionally, we flag 
the Bright Planck Multi-frequency Catalogue of Non-thermal Sources ({\bpmnt}),
 a subsample comprising 1424 bright compact sources detected with ${\rm S/N}\geq4.0$ in both the 30- and 143-GHz \planck\ channels,
described in Sect.~\ref{sec:bright}. Finally, we flag a high-significance sample of 151 objects that are
detected with the MTXF at ${\rm S/N}\geq4.0$ in {\it all\/} nine \planck\ frequency channels. We indicate this high-significance subsample with the acronym {\hpmnt}, and we discuss it in Sect.~\ref{sec:highsig}. Figure~\ref{fig:positions_PMNT} shows the Galactic coordinates of the 29\,400 candidates in our catalogue. The boundary of the \planck\ $70\,\%$ Galactic mask, \GAL070,\footnote{\GAL070\ is one of the \planck\
2015 Galactic plane masks, with no apodization used for CMB power spectrum estimation. The whole set of masks consists of \GAL020, \GAL040, \GAL060, \GAL070, \GAL080, \GAL090, \GAL097, and \GAL099, where the numbers represent the percentage of the sky that was left unmasked. These masks can be found online at the Planck Legacy Archive, \url{http://pla.esac.esa.int/pla}  See the \planck\ Explanatory Supplement for further description of the 2015 data release \citep{planck2014-ES}.} is superimposed in grey as a visual help. Apart from an evident overdensity of points around the Galactic plane and several other regions of the sky such as the Magellanic Clouds, the distribution of targets looks very uniform across the sky.

Table~\ref{tb:4sigmas} indicates the number of $4\,\sigma$ detections, that is, sources detected with ${\rm S/N}\geq4.0$,
at Galactic latitude $|b| \geq 30^{\circ}$, in the {\pmnt}, compared to the PCCS2. The $|b| \geq 30^{\circ}$
cut was set for comparison reasons, because the PCCS2 applies different Galactic masking criteria for the 
\planck\ LFI and HFI. The 
comparison is only meaningful between frequencies 30 to 143\,GHz because in the HFI channels the PCCS2
does not always apply a $4\,\sigma$ threshold cut (at LFI frequencies, on the other hand, a $4\,\sigma$ threshold 
is always applied). For HFI the situation is more complex; the lower threshold is approximately $4\,\sigma$ at 100 and
143\,GHz, $4.6\,\sigma$ at 217\,GHz, and higher at higher frequencies. For this reason and also because the
{\pmnt} is a non-blind catalogue selected at 30 and 143\,GHz, Table~\ref{tb:4sigmas} gives a fair comparison between
the two catalogues only up to 143\,GHz. The {\pmnt} catalogue presents more $4\,\sigma$ detections than the PCCS2 up
to 143\,GHz. At higher frequencies the PCCS2 has more $4\,\sigma$ detections. This is expected because the
primary selection of the {\pmnt} catalogue was performed at 30 and 143\,GHz and, therefore, we are missing the
bulk of thermal-spectrum objects that dominate the source number counts at the higher HFI frequencies. 

The gain in
number of detections with respect to the PCCS2 is particularly evident at 30, 44 and 70\,GHz. This is directly
related to the gain in S/N due to the MTXF filtering method,
but it is also the result of the de-blending
discussed in Sect.~\ref{sec:blend}.
Our primary sample has
been constructed by selecting source candidates in two \planck\ channels with different angular resolution. The
FWHM at 143\,GHz is $7\parcm3$, whereas it is $32\parcm3$ at 30\,GHz. 
As noted in Sect.~\ref{sec:blend}, it is possible that two or more distinct sources detected at 143\,GHz, with an angular 
separation smaller than $30\arcm$ could be seen as a single bright spot at 30\,GHz. 
However, since the effective angular resolution of our {\pic} is equal to the \planck\ FWHM at 143\,GHz, 
we are able to separate
sources at low radio frequencies that would not be resolved in a single frequency catalogue.
It is important to emphasize that single-frequency catalogues (such as the PCCS and PCCS2) {\it
cannot\/} resolve sources below the angular resolution limit of each separate channel, whereas the {\pmnt} is able
to resolve sources at the 143-GHz angular resolution limit, even for the LFI channels.
A more direct comparison between the two catalogues is shown in Table~\ref{tb:sources_1Jy}, where the number of detections above 1\,Jy and with Galactic latitude $|b| \geq 30\deg$ are listed for the two catalogues. Up to 143\,GHz the two catalogues contain essentially the same bright sources (the small discrepancies in the numbers of sources are compatible with the statistical random fluctuations that appear between catalogues with different flux density error levels). 
For higher frequencies, the comparison is not meaningful because the {\pmnt} is a non-blind catalogue that focuses only on the positions of our 30--143\,GHz input sample.

\begin{table}[htbp!]
\newdimen\tblskip \tblskip=5pt
\caption[]{Number of detections at Galactic latitude $|b| \geq 30^{\circ}$ in the {\pmnt} (a $4\,\sigma$ threshold has been applied) and PCCS2 catalogues (the lower threshold in the PCCS2 is $4\,\sigma$ for the channels between 30 and 143\,GHz, and higher for the channels between 217 and 857\,GHz).}
\label{tb:4sigmas}
\vskip -3mm
\footnotesize
\setbox\tablebox=\vbox{
\newdimen\digitwidth
\setbox0=\hbox{\rm 0}
\digitwidth=\wd0
\catcode`*=\active
\def*{\kern\digitwidth}
\newdimen\signwidth
\setbox0=\hbox{+}
\signwidth=\wd0
\catcode`!=\active
\def!{\kern\signwidth}
\newdimen\pointwidth
\setbox0=\hbox{.}
\pointwidth=\wd0
\catcode`?=\active
\def?{\kern\pointwidth}
\halign{\hbox to 2.0cm{#\leaderfil}\tabskip 1em&
\hfil#\hfil& \hfil#\hfil\tabskip 0em\cr
\noalign{\doubleline}
\noalign{\vskip -1pt}
\omit\hfil$\nu$ [GHz]\hfil& {\pmnt}& PCCS2\cr
\noalign{\vskip 3pt\hrule\vskip 5pt}
*30& 1243& *745\cr
*44& *814& *367\cr
*70& *959& *504\cr
100& 1440& *986\cr
143& 1430& 1283\cr
217& 1138& 1591\cr
353& *653& 1195\cr
545& *496& 1482\cr
857& *531& 4253\cr
\noalign{\vskip 3pt\hrule\vskip 5pt}}}
\endPlancktable
\end{table}

\begin{table}[htbp!]
\newdimen\tblskip \tblskip=5pt
\caption[]{Number of detections with flux density $S\geq1\,$Jy, at Galactic latitude $|b| \geq 30^{\circ}$, in the {\pmnt} and PCCS2 catalogues. }
\label{tb:sources_1Jy}
\vskip -3mm
\footnotesize
\setbox\tablebox=\vbox{
\newdimen\digitwidth
\setbox0=\hbox{\rm 0}
\digitwidth=\wd0
\catcode`*=\active
\def*{\kern\digitwidth}
\newdimen\signwidth
\setbox0=\hbox{+}
\signwidth=\wd0
\catcode`!=\active
\def!{\kern\signwidth}
\newdimen\pointwidth
\setbox0=\hbox{.}
\pointwidth=\wd0
\catcode`?=\active
\def?{\kern\pointwidth}
\halign{\hbox to 2.0cm{#\leaderfil}\tabskip 1em&
\hfil#\hfil& \hfil#\hfil\tabskip 0em\cr
\noalign{\doubleline}
\noalign{\vskip -1pt}
\omit\hfil$\nu$ [GHz]\hfil& {\pmnt}& PCCS2\cr
\noalign{\vskip 3pt\hrule\vskip 5pt}
*30& *169& *151\cr
*44& *152& *143\cr
*70& *105& *105\cr
100& **86& **82\cr
143& **76& **71\cr
217& **83& **55\cr
353& *164& **82\cr
545& *304& *570\cr
857& 1069& 2685\cr
\noalign{\vskip 3pt\hrule\vskip 5pt}}}
\endPlancktable
\end{table}

\subsection{The Bright Planck Multi-frequency Catalogue of Non-thermal Sources} \label{sec:bright}

The whole {\pmnt} catalogue contains as many entries (29\,400) as the
input \planck\ sample described in Sect.~\ref{sec:cat0}.
Even with the multi-frequency filtering many of these targets are detected with a low S/N in the LFI channels.
We therefore define a Bright Planck Multi-frequency Catalogue of Non-thermal Sources ({\bpmnt}),
containing 
those sources that have ${\rm S/N}\geq 4.0$ at both 30 and 143\,GHz simultaneously. This criterion guarantees that the members of the resulting subsample will show strong radio emission. It also favours the selection of flat radio sources. There are 1424 sources in the {\bpmnt}, among which 1146 are located outside the \planck\ \GAL070\ Galactic mask.
As will be discussed in Sect.~\ref{sec:BZCAT}, many of the sources outside the Galactic mask in the {\bpmnt} are identified as bright blazars.

\subsection{The high-significance subsample} \label{sec:highsig}

There are 151 sources in our catalogue that are detected at $\mathrm{S/N}\geq 4$ for all nine \planck\ frequencies simultaneously. We call this the high-significance subsample ({\hpmnt}).
The extragalactic sources are either bright sources with a flat spectrum continuing up to very high frequencies (normally blazars and FSRQs), or local galaxies that show both synchrotron and thermal emission, or sometimes Galactic sources that are bright enough to be detected across the entire \planck\ frequency range.

\begin{figure*}[h]
  \centering
   \includegraphics[width=\textwidth]{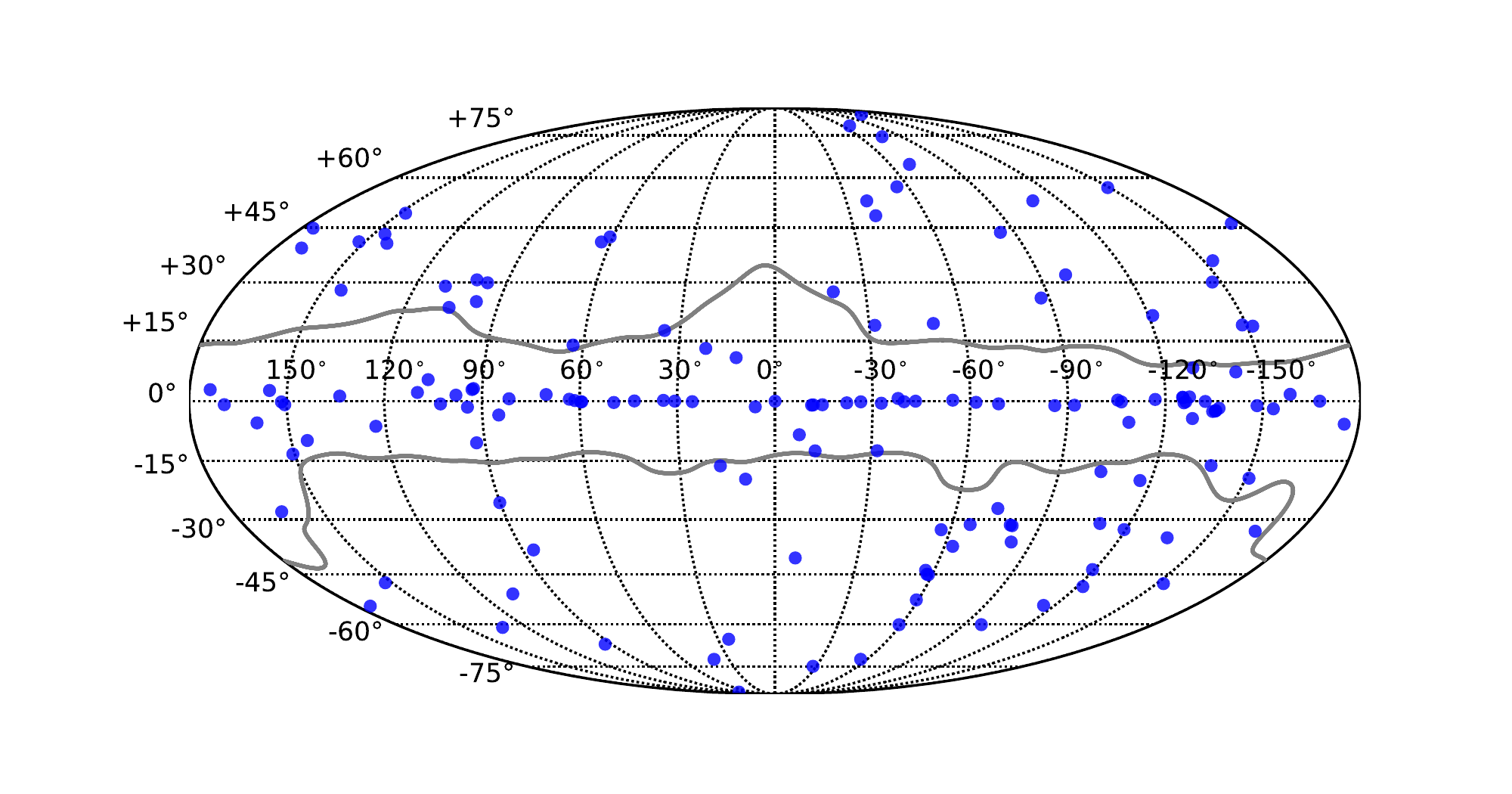}
   \caption{Position on the sky (Galactic coordinates) of the {\hpmnt} sources. The boundary of the \planck\ $70\,\%$ Galactic mask, \GAL070, is superimposed in grey. }
              \label{fig:positions_HSS}
\end{figure*}

The positions on the sky of the {\hpmnt} sources are shown in Fig.~\ref{fig:positions_HSS}.
As can be seen from the figure, many {\hpmnt} sources (76 out of 151) lie inside the \planck\ \GAL070\ Galactic mask. This was expected, since the criterion of setting a $4\,\sigma$ threshold at all frequencies selects the brightest sources on the sky, and most of these ultra-bright sources lie in the Galactic plane.

\section{Validation and properties of the catalogues} \label{sec:valprop}

\subsection{Internal \planck\ validation} \label{sec:fluxval}

In order to validate the flux densities obtained with the MTXF multi-frequency detection
technique we have compared the
{\pmnt} flux densities
 with those of the PCCS2 \citep{planck2014-a35}. The PCCS2 lists four different flux density estimates for each of its sources: the native detection method flux estimation (\texttt{DETFLUX}); aperture photometry (\texttt{APERFLUX}); Gaussian fitting (\texttt{GAUFLUX}); and point-spread function fitting (\texttt{PSFFLUX}). Except for the case of very extended sources, the four estimations tend to show good agreement. Here we choose the \texttt{DETFLUX} flux density estimation as a reference.

\begin{figure*}[h]
   \centering
   \includegraphics[width=\textwidth]{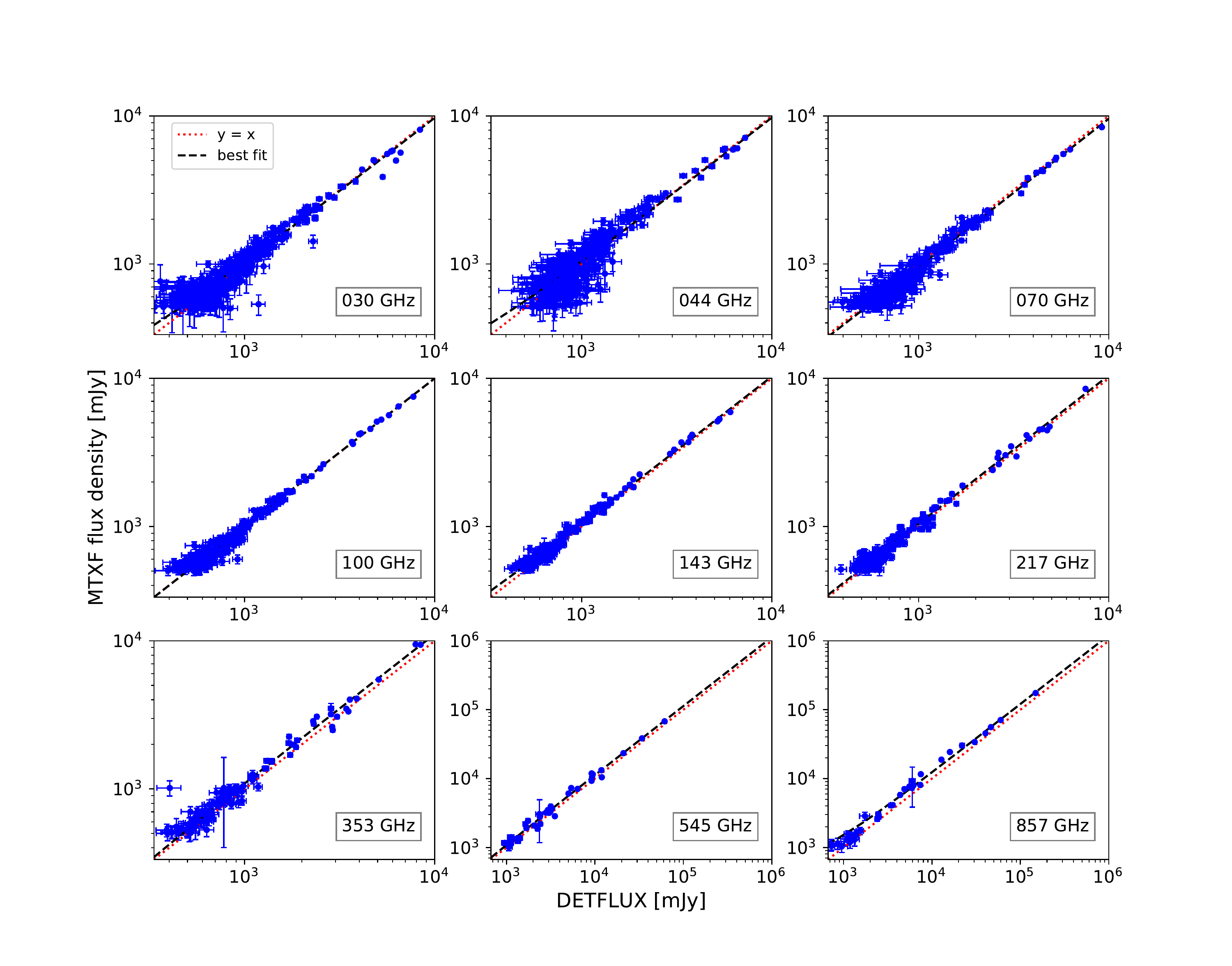}\\
   \caption{{\pmnt} flux densities versus PCCS2 \texttt{DETFLUX} at the same frequency. The $x=y$ identity is shown with a red dotted line. The dashed line shows the best linear fit to the data points and their error bars for sources with PCCS2 \texttt{DETFLUX} between 0.5 and 10\,Jy up to 353\,GHz and above 1\,Jy at higher frequencies.}
              \label{fig:val_all}%
    \end{figure*}

Our input catalogue contains 29\,400 sources, many of them not included in the PCCS2. For this validation step we select for each frequency only sources outside the \planck\ \GAL040\ Galactic mask. This is a more restrictive mask than the one we use in the rest of this paper (GAL070); the rationale for this choice is that we want to be sure to avoid Galactic sources as much as possible for this comparison.  There are two reasons for this.  Firstly,
Galactic sources will be seen against a brighter background, which could cause errors in the determination of their flux densities; secondly, a significant portion of Galactic objects appear as extended sources, particularly for the HFI channels. As discussed in Sect.~\ref{sec:photometry}, accurate filter photometry 
requires that the sources are point-like (i.e. smaller than the beam area). 
Figure~\ref{fig:val_all} shows results of the comparison between MTFX and PCCS2 flux densities for the nine \planck\ channels. In order to quantify the degree of agreement between {\pmnt} and PCCS2 flux density estimates, we have performed a linear fit based on the orthogonal distance regression method \citep{boggs90}, which takes into account uncertainties in both the $x$ and $y$ axes.\footnote{The number of points that satisfy our criteria and are used for the fitting at 
30, 44, 70, 100, 143, 217, 353, 545, and 857\,GHz are
354, 232, 217, 198, 159, 122, 96, 34, and 36, respectively.}
The resulting coefficients to the fit
$S_\mathrm{{\pmnt}} = a \, S_\mathrm{PCCS2}+b$ are shown in Table~\ref{tb:fits}. For this fit only the sources with $0.5\,{\rm Jy}<S_\mathrm{PCCS2}<10{\rm Jy}$ were used, except for the cases of 545 and 857\,GHz, where the Galactic contamination is stronger. For these two channels, we have just considered sources above 1\,Jy.

{\pmnt} and PCCS2 flux densities agree very well for the channels up to 217\,GHz. At 353\,GHz and above the {\pmnt} tends to overestimate flux densities with respect to PCCS2. There are two possible reasons for this disagreement. 

The first reason is background contamination: because the HFI channels are more contaminated by Galactic and extragalactic infrared emission, and since the MTXF approach picks up the maximum of the filtered images around the positions of each target in the input catalogue, flux density estimates can be overestimated in regions with strong contamination. Galactic emission at HFI frequencies, particularly at 545 and 857\,GHz, is strong even outside the Galactic mask. 
For those frequencies, other flux density estimators (such as aperture photometry) can be more robust.
Our choice of the restrictive 
\GAL040\ mask, which leaves only $40\,\%$ of the sky unmasked, and the relatively high threshold of 1\,Jy we set for the 545- and 857-GHz channels, should largely reduce the effect of this source of overestimation. 

The second reason for the disagreement between \pmnt\ PCCS2 flux densities is a fundamental limitation of the multi-frequency filtering technique, 
 discussed in Appendix~\ref{sec:practical}. The MTXF technique mixes 
data from different channels according to Eq.~(\ref{eq:matrix_filters_eqs}). This could lead to intensity leakage between channels, but this is prevented by the orthornomality condition (Eq.~\ref{eq:orthonormal}). The problem is that Eq.~(\ref{eq:orthonormal}) works only if the source profiles $\tau_l (\vec{q})$ are well known for all the frequencies about to be filtered. For this paper we have assumed circular Gaussian beams with the \planck\ nominal FWHM values. This assumption is good enough if: (a) sources are point-like; and (b) the instrumental beams are well described by a circular Gaussian (i.e. the beam is stable across the image, with small sidelobes, and circular symmetry). This assumption is not correct in two relevant cases:
\begin{itemize}
\item when beams are significantly non-Gaussian and non-circular, as it is discussed in Appendix~\ref{sec:practical};
\item for sources in the high-frequency \planck\ channels that are extended, which is a problem, since if a source is not strictly point-like, the orthonormality relation (Eq.~\ref{eq:orthonormal}) is not satisfied and there will be some leakage between channels, affecting the MTXF photometry.
\end{itemize}
In order to check this second point we have repeated the fit but using only sources that are not flagged as extended in PCCS2. Below 217\,GHz the effect of excluding extended sources in the fit is negligible. Table~\ref{tb:fits2} shows the fits for 217, 353, 545, and 857\,GHz after excluding the sources that are flagged as extended in the PCCS2. If we do not consider the fit errors, the fit is slightly better at 217\,GHz, significantly better at 353\,GHz, and only marginally better at 545 and 857\,GHz, where the effect of Galactic contamination combined with pointing inaccuracies dominates the photometric errors.\footnote{A fit at 857\,GHz between PCCS2 and {\pmnt} flux densities (using the MTXF flux density estimation at the pixel corresponding to
the
exact coordinates in the input sample instead of the local maximum around that position) gives fit parameters $a = 0.94 \pm 0.04$ and $b = (-350 \pm 485)\,$mJy. Excluding sources flagged as extended, we get $a = 0.90 \pm 0.05$ and $b = (27 \pm 1168)\,$mJy. Now most of the bias due to random positive background fluctuations is removed, but we underestimate fluxes because the coordinates in the input sample (obtained at 30 and 143\,GHz) do not necessarily correspond to the true positions at 857\,GHz.} Taking into account the errors in the fits, the only channel for which it is important to remove the sources flagged as extended is 353\,GHz.

\begin{table}[htbp!]
\newdimen\tblskip \tblskip=5pt
\caption[]{Results of the fit $S_\mathrm{MTXF} = a S_\mathrm{PCCS2}+b$ for the nine \planck\ frequencies.}
\label{tb:fits}
\vskip -3mm
\footnotesize
\setbox\tablebox=\vbox{
\newdimen\digitwidth
\setbox0=\hbox{\rm 0}
\digitwidth=\wd0
\catcode`*=\active
\def*{\kern\digitwidth}
\newdimen\signwidth
\setbox0=\hbox{+}
\signwidth=\wd0
\catcode`!=\active
\def!{\kern\signwidth}
\newdimen\pointwidth
\setbox0=\hbox{.}
\pointwidth=\wd0
\catcode`?=\active
\def?{\kern\pointwidth}
\halign{\hbox to 2.0cm{#\leaderfil}\tabskip 1em&
\hfil#\hfil& \hfil#\hfil\tabskip 0em\cr
\noalign{\doubleline}
\noalign{\vskip -1pt}
\omit\hfil$\nu$ [GHz]\hfil& $a$& *!$b$ [mJy]\cr
\noalign{\vskip 3pt\hrule\vskip 5pt}
*30& $0.97\pm0.01$& $*!65.3\pm*12.0$\cr
*44& $0.97\pm0.01$& $*!75.9\pm*25.8$\cr
*70& $0.96\pm0.01$& $**!3.3\pm*10.8$\cr
100& $1.00\pm0.01$& $**-1.4\pm**8.0$\cr
143& $1.02\pm0.01$& $*!26.7\pm**8.2$\cr
217& $1.05\pm0.01$& $**-4.6\pm*13.8$\cr
353& $1.11\pm0.02$& $*-24.7\pm*25.2$\cr
545& $1.13\pm0.02$& $*-40.7\pm150.3$\cr
857& $1.21\pm0.02$& $!295.3\pm234.9$\cr
\noalign{\vskip 3pt\hrule\vskip 5pt}}}
\endPlancktable
\end{table}

\begin{table}[htbp!]
\newdimen\tblskip \tblskip=5pt
\caption[]{Results of the fit $S_\mathrm{MTXF} = a S_\mathrm{PCCS2}+b$ for the four highest \planck\ frequencies, excluding sources that are flagged as extended in the PCCS2.}
\label{tb:fits2}
\vskip -3mm
\footnotesize
\setbox\tablebox=\vbox{
\newdimen\digitwidth
\setbox0=\hbox{\rm 0}
\digitwidth=\wd0
\catcode`*=\active
\def*{\kern\digitwidth}
\newdimen\signwidth
\setbox0=\hbox{+}
\signwidth=\wd0
\catcode`!=\active
\def!{\kern\signwidth}
\newdimen\pointwidth
\setbox0=\hbox{.}
\pointwidth=\wd0
\catcode`?=\active
\def?{\kern\pointwidth}
\halign{\hbox to 2.0cm{#\leaderfil}\tabskip 1em&
\hfil#\hfil& \hfil#\hfil\tabskip 0em\cr
\noalign{\doubleline}
\noalign{\vskip -1pt}
\omit\hfil$\nu$ [GHz]\hfil& $a$& *!$b$ [mJy]\cr
\noalign{\vskip 3pt\hrule\vskip 5pt}
217& $1.04\pm0.04$& $!**4.9\pm*52.1$\cr
353& $0.98\pm0.04$& $!*91.0\pm*94.7$\cr
545& $1.13\pm0.04$& $-268.6\pm324.0$\cr
857& $1.20\pm0.05$& $!391.9\pm849.4$\cr
\noalign{\vskip 3pt\hrule\vskip 5pt}}}
\endPlancktable
\end{table}

\subsection{External validation} \label{sec:ext_val}

The {\pmnt} contains 29\,400 entries, so it is not practical to perform an individual external validation of all the candidates. Instead of following such an approach, we attempt 
statistical external validation by comparing the number counts of our catalogue with existing models. We also try to validate a subset of interesting bright {\pmnt} sources that are not present in the PCCS2 by looking for matches in existing ground-based observing databases.
Additionally, 
we study the reliability and completeness of the catalogue by comparison to 
the Combined Radio All-Sky Targeted Eight-GHz Survey catalogue \citep[CRATES,][]{CRATES}
and
we cross-correlate the non-thermal-dominated {\bpmnt} and {\hpmnt} catalogues with a catalogue of known blazars.

\subsubsection{Number counts} \label{sec:numbercounts}

One simple approach to check the validity of the {\pmnt} catalogue is to estimate its number counts and to
compare them with well known models and/or other data.
For this purpose we have considered only extragalactic (outside the \GAL070\ \planck\ Galactic mask)
sources detected at the $>4\,\sigma$ level in each channel. Another complementary approach
is to study the distributions of spectral indices of the populations in the catalogue, as well as their
variation with frequency, and to compare these with models and with previous observations.

\begin{figure}[t]
  \centering
   \includegraphics[width=\columnwidth]{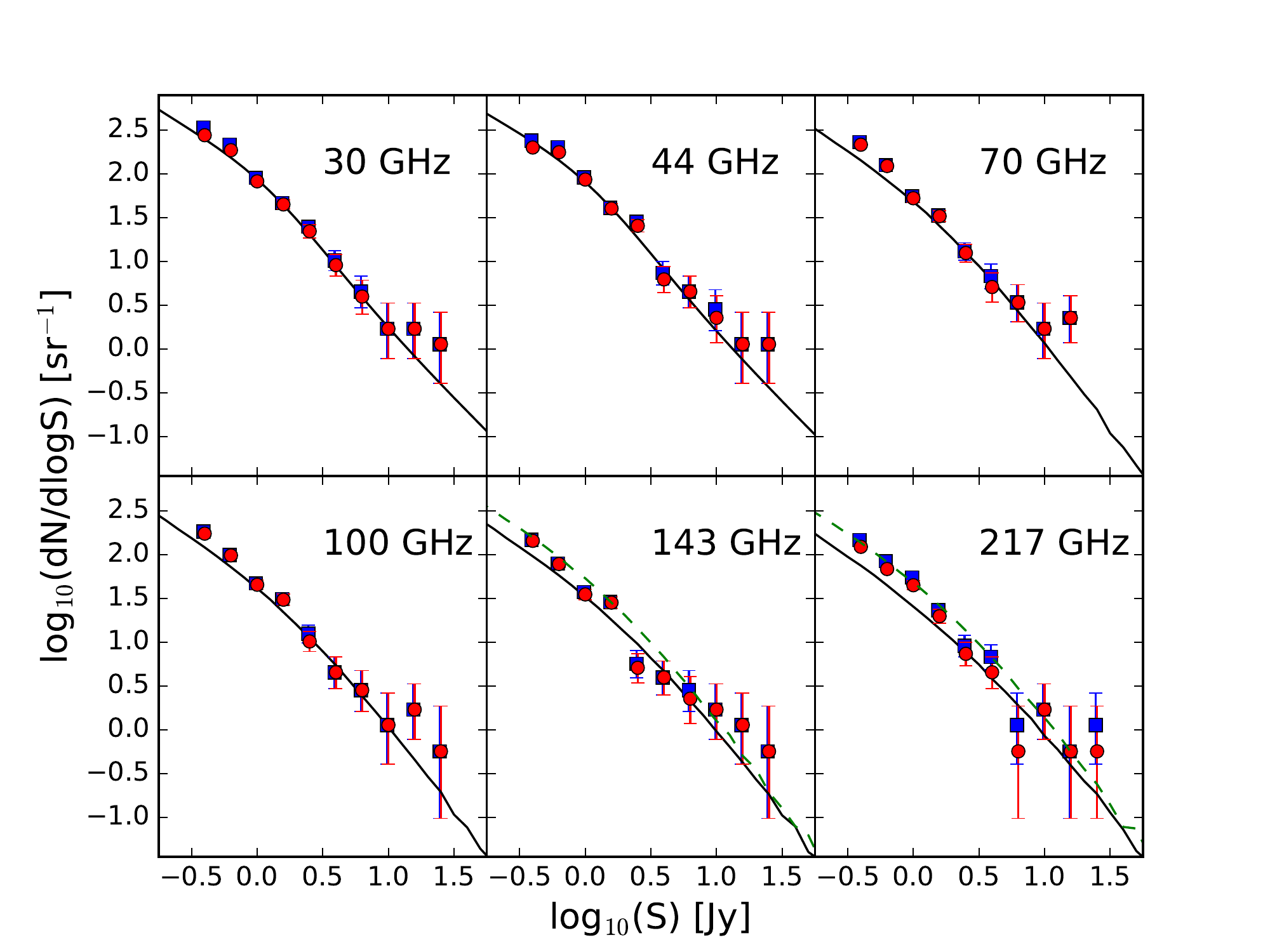}
\caption{{\pmnt} differential source number counts for 30--217\,GHz here. Only $4\,\sigma$ detected sources at each frequency outside the \GAL070\ \planck\ Galactic mask were used (blue squares). The red
circles correspond to the source counts estimated only for sources with a low-frequency counterpart in the CRATES \citep{CRATES} catalogue. As a comparison, the predicted radio source number counts from the
\cite{tucci11} C2Ex model
are also plotted (continuous black line).
In the case of the 143 and 217 GHz channels, we also plot the number counts predicted by 
the \cite{tucci11} C2Co model (dashed green line). \label{fig:counts_all} }
\end{figure}

Figure~\ref{fig:counts_all} displays the {\pmnt} differential source number counts (blue dots) in the six \planck\ channels between 30 and 217 GHz.  Only sources above the $4\,\sigma$ detection threshold and outside the \GAL070\ Galactic mask are considered. The corresponding numbers of sources -- used at each frequency channel -- are listed in Table~\ref{tb:summary} (first row). As a check, we also plot the differential number counts (red dots) calculated by using the same source sample, but selecting only those sources with a low-frequency counterpart in the CRATES \citep{CRATES} catalogue (see Section~\ref{sec:ext_val}). Both estimates appear in very good agreement at each \planck\ frequency channel and in the whole flux density interval here analysed, thus showing that the source selection we adopted misses only a negligible fraction of low-frequency radio sources.

As a comparison, in each panel we plot the differential source number counts predicted by the C2Ex model of \cite{tucci11}. This was the most successful of their models for predicting number counts at high radio frequencies (i.e. $\nu$> 100 GHz), given that it consistently explained various and independent data sets 
\citep{Vieira10,Marriage11,planck2011-6.1}, in the flux density interval $0.05< S <10$ Jy. In the case of the 143 and 217 GHz channels, we also plot the number counts predicted by the C2Co model discussed by \cite{tucci11}.\footnote{Moving to lower \planck\ frequencies, the differences between the models C2Ex and C2Co gets smaller and smaller; in particular, their predictions are very similar at 30 and 44 GHz.  For this reason, we did not plot the C2Co model at lower frequencies.} This latter model assumed a more compact synchrotron-emission region -- in the inner jet of FSRQs -- than C2Ex, while keeping the same parameters for BL Lac objects
\citep[see][their Section 4.2]{tucci11}. This assumption implies that, in the C2Co model, an almost flat emission spectrum can be maintained up to higher radio frequencies, thus increasing the number of detected sources at high frequencies.

 Regarding the counts between 353 and 857\,GHz (not shown in the plot),
 we notice an excess of sources with 
 respect to the model above 300\,mJy.  Although there are partial reasons for this (for example local galaxies showing both radio and infrared emission), the simplest explanation could be contamination by thermal emission that appears in the same positions of (already fainter) radio sources detected at lower \planck\ frequencies, or purely thermal sources (either Galactic or extragalactic) that are bright enough to be selected at a frequency as low as 143\,GHz in our input sample.  
We discuss this possibility in Sect.~\ref{sec:Galactic}.

\subsubsection{Reliability and completeness} \label{sec:relcomp}

To verify the reliability of our catalogue we cross-match our detected sources with the CRATES catalogue
 \citep{CRATES} at 8.4\,GHz. However, by construction, the CRATES catalogue misses the steep-spectrum sources. Since we expect to include several sources of this type, we also complement the cross-match (within a 15\arcm\ search radius) with the GB6 
\citep[Northern hemisphere,][]{GB6}
and PMN 
\citep[Southern hemisphere,][]{PMN} catalogues, both at $5\,$GHz. Additionally, we 
also use other low-frequency catalogues and ground-based observing databases, such as
The Australia Telescope 20\,GHz (AT20G) survey \citep{AT20G} or the SIMBAD 
database.\footnote{\href{http://simbad.u-strasbg.fr/}{http://simbad.u-strasbg.fr/}}
Since the {\pmnt} extends to IR frequencies, on occasion we also use higher frequency catalogues 
such as those constructed using IRAS \citep{IRAS} or the 
Planck Catalogue of Galactic Cold Clumps \citep[PGCC,][]{planck2014-a37} in order to check 
if particular candidates with thermal-like emission can be matched to already known sources.
However, our main focus in this paper will be the reliability and completeness of the 
low-frequency, non-thermal part of the catalogue. Due to the double
{\pmnt} selection criterion at 30\,GHz and 143\,GHz, we will study the reliability of the catalogue
in consecutive steps, starting from sources detected with ${\rm S/N}\geq 4$ separately in each one of the two selection channels and then proceeding to a more restrictive, brighter subsample of 
sources detected above the $4\,\sigma$ level simultaneously at all the non-thermal \planck\ channels (30, 44, 70, 100, and 143\,GHz). Finally, we will discuss the completeness of the catalogue.

\begin{figure}[h]
 \centering
\includegraphics[width=\columnwidth]{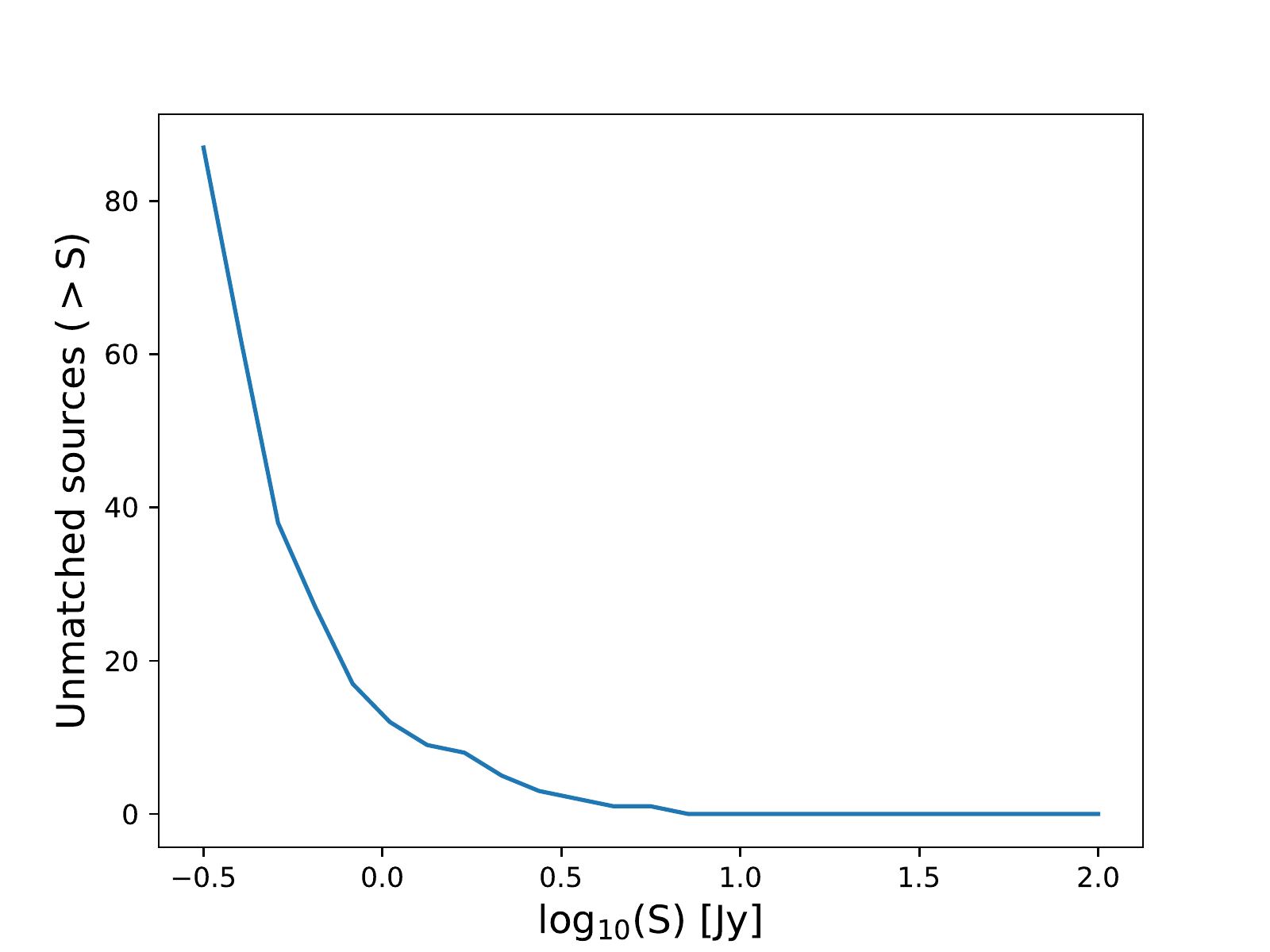}
\caption{Number of {\pmnt} sources with no low-frequency counterpart as a function of the flux density at 30\,GHz. }
\label{fig:reliab}
\end{figure}

\begin{figure}[h]
 \centering
\includegraphics[width=\columnwidth]{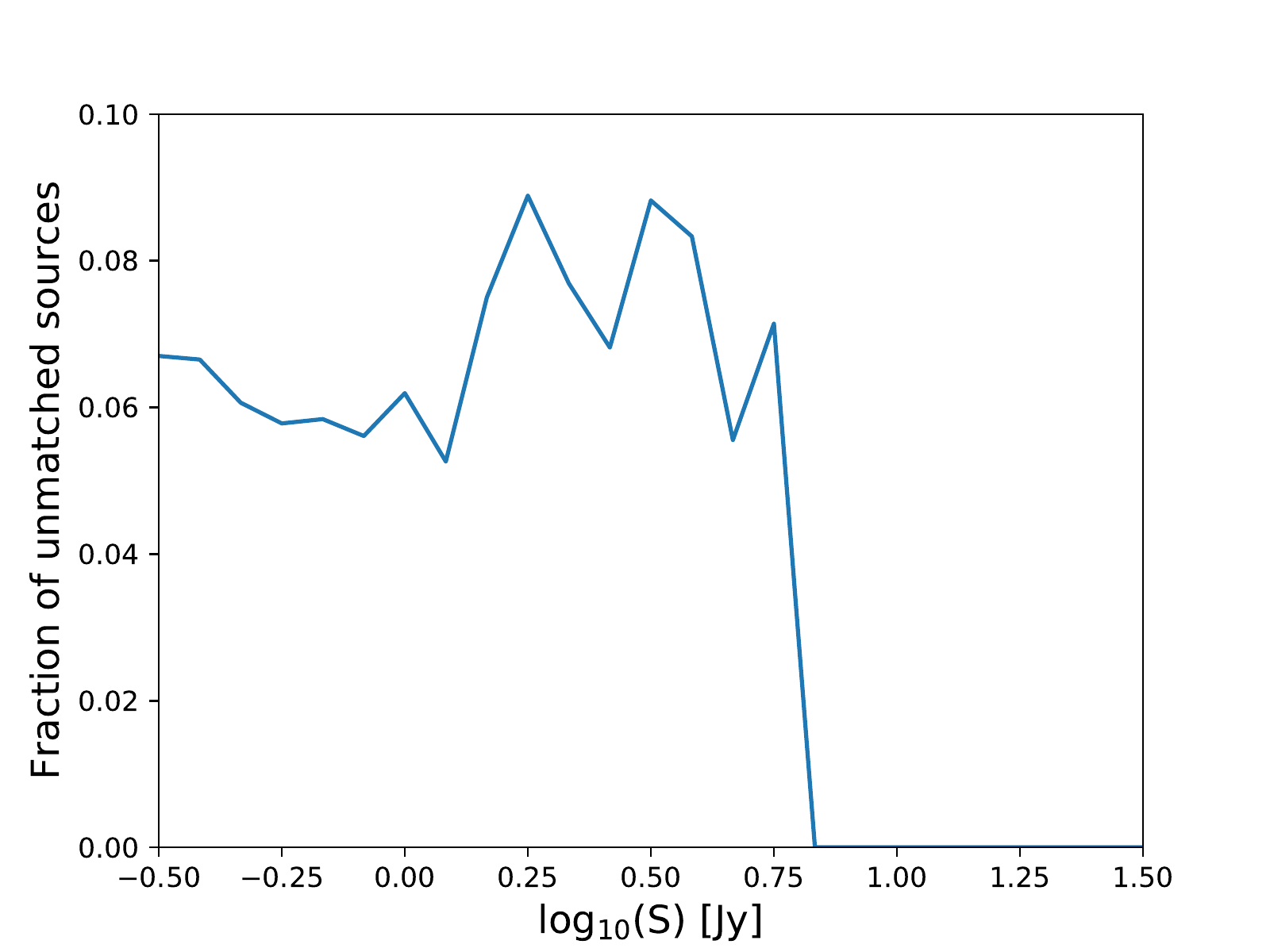}
\caption{Fraction of {\pmnt} sources with no low-frequency counterpart as a function of the flux density at 30\,GHz. }
\label{fig:frac_reliab}
\end{figure}

\noindent
\textbf{${\rm S/N}\geq4$ at 30\,GHz:} At 30\,GHz we have 1701 detected sources 
outside the Galactic mask \GAL070\ that have ${\rm S/N}\geq4$. Of these, 1566 have a counterpart in the low-frequency catalogues, implying that at most $8\,\%$ of the selected sources are spurious. In 
Fig.~\ref{fig:reliab}
we can see the number of unmatched sources as a function of the flux density,
while Fig.~\ref{fig:frac_reliab} shows
 the fraction of unmatched sources as a function of the flux density.
 As expected, most of the unmatched sources appear near the detection limit because they are mostly caused by artefacts produced by nearby bright sources (they are all flagged as part of a group). 
 
\noindent
\textbf{${\rm S/N}\geq 4$ at 143\,GHz:} The other interesting channel to check the reliability of our catalogue is at 143\,GHz, the second channel used to obtain the candidates of the initial sample. At this frequency we have 2047 sources outside the \GAL070\ \planck\ Galactic mask and with ${\rm S/N}\geq4$. There are 1664 sources with counterparts in the lower frequency radio catalogues and 1720 if we also consider IRAS \citep{IRAS} counterparts. Among the 327 
sources not matched to these catalogues, only 
14 are in the {\bpmnt} and are brighter than 200\,mJy at 143\,GHz. 
The rest have low flux densities at 143\,GHz. Returning to the 14 relatively bright {\bpmnt}
sources that are not matched to GB6 or PMN,
nine of them have a counterpart in the CRATES catalogue. Among the remaining five objects, four have cross-identification with the BZCAT5 catalogue of blazars \citep{massaro15}; they were also observed by AT20G, with flux densities at 20\,GHz that are consistent with our 30\,GHz photometry within factors of 0.9--1.2. The last {\bpmnt} object in this list is 33.7 arcseconds away
from blazar candidate WISE J140610.82$-$070702.4 \citep{WISEp}, whose flux density at 20\,GHz measured by AT20G is also
compatible with the {\pmnt} 30-GHz flux density.
Of the remaining unmatched sources that
 are not in the {\bpmnt}, 16 are brighter than 200\,mJy at 143\,GHz. However, eight of these have counterparts in the
PGCC \citep{planck2014-a37}, so probably they are associated with regions of Galactic emission. 
Of the remaining eight objects, four are detected only at 143\,GHz or at 143\,GHz and one other channel and, therefore, it is reasonable to consider them spurious detections. The remaining objects are of uncertain nature, but all of them (except for one) have flux densities below $300\,$mJy at 143\,GHz; they are also probably spurious detections. The last one, {\pmnt} ID\,4888, has a measured flux density of 482.4\,mJy at 143\,GHz and is detected above the $4\,\sigma$ level between 70 and 353\,GHz in the {\pmnt}.

\bigskip

\noindent
\textbf{${\rm S/N}\geq 4$ between 30 and 143\,GHz:} The {\pmnt} contains 1012 sources detected above the $4\,\sigma$ level, simultaneously
at 30, 44, 70, 100, and 143\,GHz. Among these 1012 sources, 19 
do not have a counterpart in PCCS2 in any
of the aforementioned frequency bands.
All of these 19 sources have counterparts within a 15\arcm\ search radius either in the GB6 or the PMN catalogues. 
In order to further confirm that these detections are real ERSs with good photometric
measurements in the {\pmnt}, 
we looked for
matches in existing ground-based observing databases, which have data
reaching to frequencies closer to the lower bands of the \planck-LFI than the GB6 and the PMN. We excluded sources close to the Galactic plane ($|b|\leq
10^{\circ}$) to avoid contamination by Galactic emission. This left
us with 18 of 19 sources.

The AT20G Survey covers the sky south of
declination $0^{\circ}$. We found 10 matches, each within 6\arcs\ to 93\arcs\ of the
{\pmnt} position, and one within 125\arcs. The 20-GHz flux densities were taken
between 2004 and 2008, the observing dates depending on the declination band. Thus
none of the observations exactly overlap with the \planck\ observations, which took
place
between 12 August 2009 and 23 October 2013. The AT20G
20-GHz flux densities agree with the {\pmnt} 30-GHz flux densities within a factor of 0.6 to 2.7 (average ratio 1.06).  This
is in good agreement with what is expected for possibly variable sources when
observations are taken at two different centimetre-domain frequency bands and at two
different observing epochs several years apart \citep{hovatta07,nieppola07}.

For sources with a declination higher than $-20\deg$, we checked the Owens
Valley Radio Observatory \citep[OVRO,][]{OVRO} 40-m telescope database \citep{OVROpaper}.\footnote{\url{http://www.astro.caltech.edu/ovroblazars/}}
Starting in 2008, they
regularly monitored 1800 blazars at 15\,GHz. We found four matches within 86\arcs\
or less of our new {\pmnt} source positions; one of them was also in our AT20G source
identification list. All of these sources showed at least some 15-GHz
variability during the \planck\ mission.  For all of these four sources the
30-GHz {\pmnt} flux densities are within a factor of 2 or less of their 15-GHz
long-term average flux densities, thus, considering their variable behaviour, the flux densities
are of comparable amplitudes.

Visual inspection at 30 and 143\,GHz of the \planck\ maps around the three remaining $4\,\sigma$ candidates (that are not matched either to AT20G or OVRO sources) suggests the presence of moderate point-like, positive
temperature fluctuations at the positions of three of the targets (ID numbers $2037$, $14439$, and $2140$), but does not reveal any obvious structure around the remaining
target (ID\,$28805$). These four sources, however, all have
 counterparts at 5\,GHz in the GB6 or the PMN, and their
 {\pmnt} flux densities at 30\,GHz are compatible with those of their low-frequency counterparts, strongly suggesting that 
 these objects are real flat spectrum radio sources.

\bigskip
\noindent
\textbf{Completeness:} From the source number counts (see Sect.~\ref{sec:numbercounts}) we can also derive a rough estimate of the completeness limit of the catalogue. This lies at around 300\,mJy up to 70\,GHz, decreasing to about 150\,mJy out to 217\,GHz. At higher frequencies it continuously increases, with 250, 500 and 1000\,mJy limits for 353, 545, and 857\,GHz, respectively.

\subsection{Cross-identification of {\bpmnt} sources with the BZCAT5 catalogue of blazars} \label{sec:BZCAT}

As noted in Sects.~\ref{sec:cat0} and~\ref{sec:cat1}, the criteria we 
set for the creation of our input catalogue favours the selection of non-thermal compact sources. This is
particularly true for the {\bpmnt} and the {\hpmnt}, which should by construction contain
many bright flat-spectrum sources. In order to check this, we have cross-correlated both sub-catalogues with an external catalogue of known blazars.
913 out of the 1424 sources of the
{\bpmnt} have counterparts within $7^{\prime}$ (the \planck\ FWHM at 143\,GHz) in the BZCAT5 catalogue of blazars \citep{massaro15}. Most of them, 832, are outside the \planck\ \GAL070\ Galactic mask.
The BZCAT5 provides flux density estimation at 143\,GHz from the PCCS catalogue \citep{planck2013-p05}
for 472 of the 913 blazars with counterpart in the {\bpmnt}.
Remarkably, thanks to our
multi-frequency MTXF photometry, the {\bpmnt} can provide flux density estimations at 143\,GHz and the corresponding photometric errors for the other 441 blazars in the list. 

Among the 151 sources in the {\hpmnt}, 72 have a
counterpart within a radius of 7\arcm\ in the BZCAT5 catalogue. 60 of these objects
are outside the \planck\ \GAL070\ Galactic mask, and 49 of these have flux densities at 143\,GHz listed in
the BZCAT5. Thanks to our
multi-frequency MTXF photometry, we can provide the flux densities at 143\,GHz for the 
remaining 11 BZCAT5 objects.

\begin{figure}[h]
  \centering
   \includegraphics[width=\columnwidth]{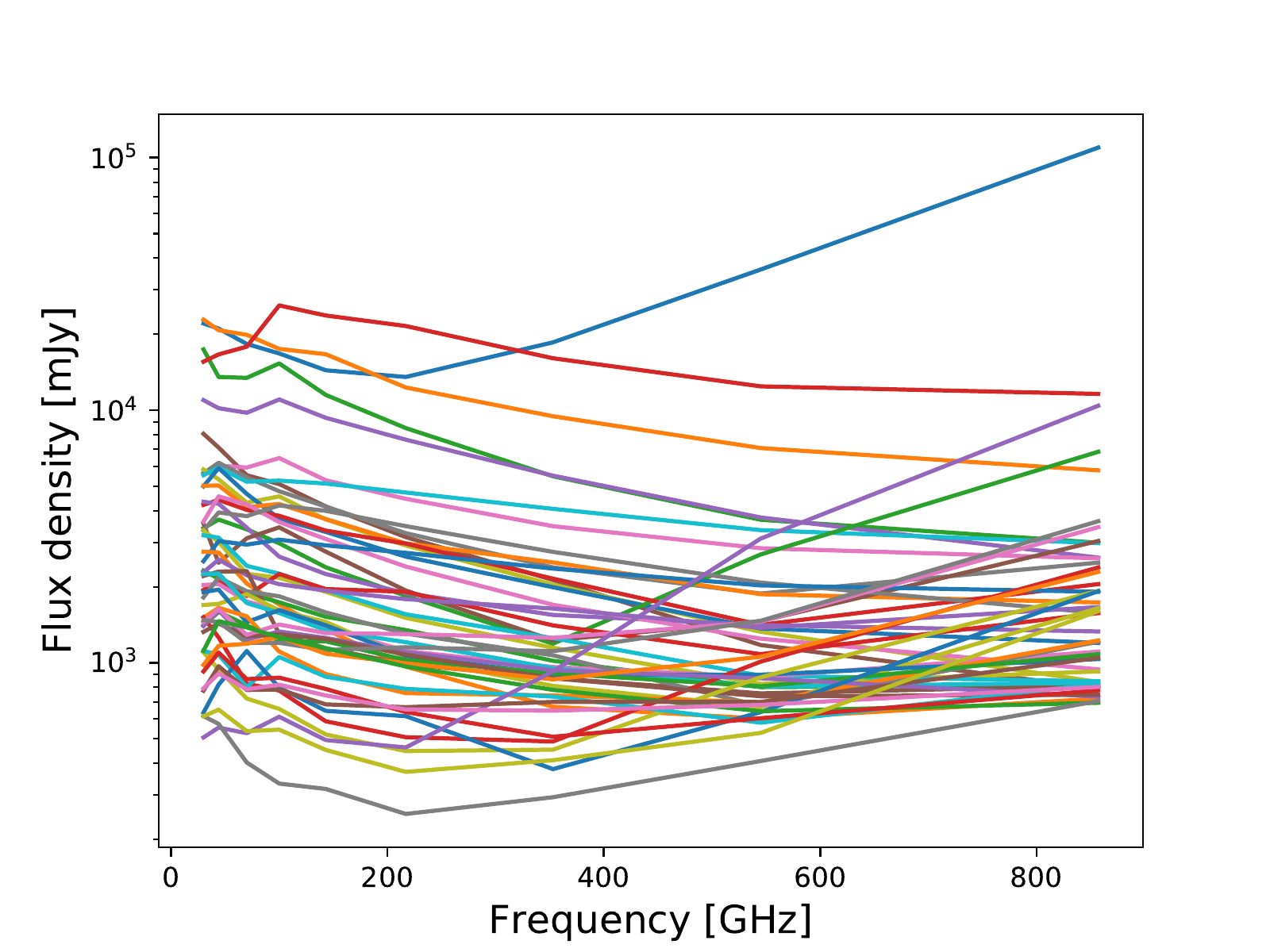}
   \caption{Spectral energy distributions of the 49 {\hpmnt} sources outside the \planck\ \GAL070\ mask that are identified as blazars in the BZCAT5 catalogue \citep{massaro15}.}
              \label{fig:HSS_massaro_fluxes}
\end{figure}

\begin{figure}[h]
  \centering
   \includegraphics[width=\columnwidth]{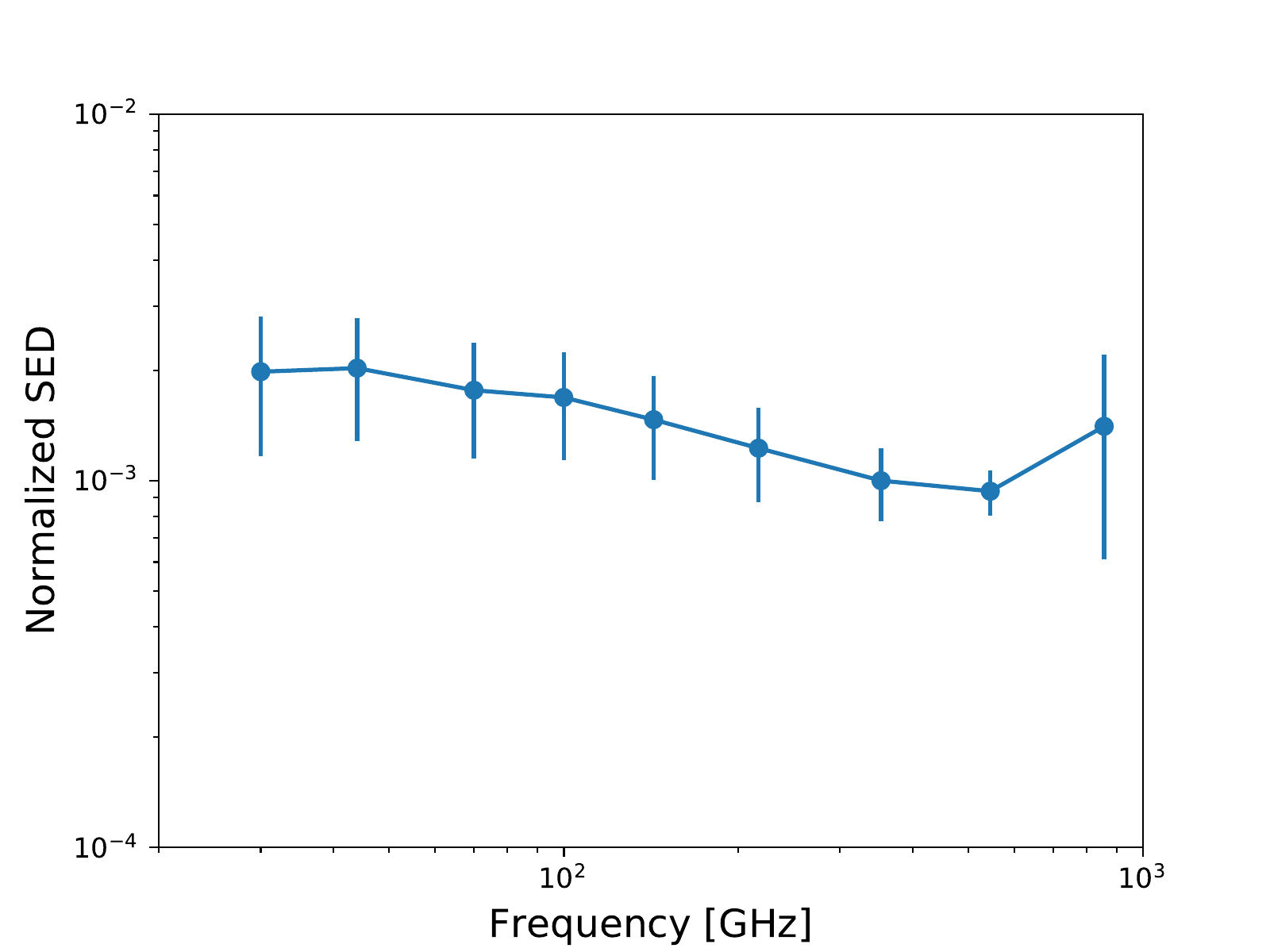}
   \caption{Average normalized spectral energy distributions of the 49 {\hpmnt} sources outside the \planck\ \GAL070\ mask that are identified as blazars in the BZCAT5 catalogue \citep{massaro15}.}
              \label{fig:average_SEDS}
\end{figure}

Figure~\ref{fig:HSS_massaro_fluxes} shows the {\pmnt} SEDs of the 49 sources of our high-significance subsample
that have flux densities at 143\,GHz listed in
the BZCAT5.\footnote{The BZCAT5 catalogue provides flux density information at 143\,GHz for only $16\,\%$ of its sources.}  Not
surprisingly, almost all these blazar SEDs are essentially flat in the whole frequency range analysed here. 
Figure~\ref{fig:average_SEDS} shows the average normalized SED of the same 49 sources. This SED is essentially flat, with an excess at 857\,GHz that is partially due to contamination from Galactic emission and, very occasionally, to random associations with nearby galaxies.
There are at least two sources whose
SEDs rise sharply at high frequency. The first one is the brightest object in the plot, which is
only 1\arcm\ away from the NASA/IPAC Extragalactic Database (NED\footnote{The NASA/IPAC Extragalactic Database (NED) is operated by the Jet Propulsion Laboratory, California Institute of Technology, under contract with the National Aeronautics and Space Administration.}) coordinates of the galaxy 
Centaurus~A. The host galaxy of Cen~A has a dust lane,
responsible for powerful FIR emission. The Cen~A \emph{Herschel} 
flux densities at 500 and 350\,$\mu$m are roughly 35 and 100\,Jy \citep[as obtained by visual inspection of the right panel of figure~3 in][]{Parkin}, 
to be compared to our flux densities of 36 and 110\,Jy at 545 and 857\,GHz, respectively. Regarding the second object,\footnote{The BZCAT5 redshift for this source is $z=0.7$} whose
flux density at 857\,GHz rises to become the third highest in the plot (light purple line in Fig.~\ref{fig:HSS_massaro_fluxes}), it is located 5\arcm.4 away from nearby galaxy NGC 6503 
\citep[$z = 0.000083$,][]{Epinat08}. This random association could explain the observed flux excess at high frequencies for this object.

Many of the SEDs in Fig.~\ref{fig:HSS_massaro_fluxes} show a small but noticeable bump around 100\,GHz. A possible explanation for this excess could be the signature of CO contamination, which presents its strongest transition line ($J\,{=}\,1\,{\rightarrow}\,0$) at 115
GHz, very close to this frequency channel. This possible CO excess could also have 
an extragalactic origin, since the redshifted $J\,{=}\,1\,{\rightarrow}\,0$ transition should be observable in the \planck\ 100-GHz band for sources with $z \lesssim 0.45$, and this criterion is met by 19 out of the 49 sources plotted in Fig.~\ref{fig:HSS_massaro_fluxes}. 
However, the CO emission from blazars is generally considered to be negligible. We have tested this by exploiting the fact that
there is a tight, almost linear, relationship between the CO and the total IR luminosity due to dust emission \citep[e.g.][]{Greve14}. Therefore is possible to estimate the total IR luminosity of a source from its observed excess at 100\,GHz, assuming that such excess is purely due to CO emission, and to compare to direct observations, where available. When IR observations are not available, it still possible to compute the star formation rate (SFR) from the IR emission derived from the CO excess, using the calibration of \cite{Kennicutt12}, and check if the derived S/N is plausible. This kind of comparison is only approximate, but can give us an idea of whether the observed excess could be due to extragalactic CO emission. Only in the case of Cen~A is the IR luminosity from the CO consistent with such direct measurements. In two other cases, {\pmnt} ID\,721 and ID\,1257, the derived SFRs are plausible (500--600$\,{\rm M}_{\odot}\,{\rm yr}^{-1}$); however, there is no indication that either of these sources have any significant star formation (and low-$z$ blazars do not generally show significant SFRs). The CO excesses of all the other sources would correspond to IR luminosities orders of magnitude higher than observed, or to SFRs far above the Eddington limit (${\rm SFR}\gg10^4\,{\rm M}_{\odot}\,{\rm yr}^{-1}$). We conclude that, with only a few exceptions (most notably Cen~A), the CO emission from blazars in our sample is unimportant. 

A second possibility is that the observed excess is caused by Galactic CO emission, visible at high Galactic latitudes \citep[cf.][figures 3, 4, and 5, and the corresponding discussion in section~5 of that paper]{Planck2013-p13}. If that were the case, the 100-GHz excess would also be visible in other {\pmnt} sources at similar Galactic latitudes and not in the BZCAT5. While it is true that the excess is visible for some of the {\pmnt} sources not in the BZCAT5, there is no noticeable bump in the averaged SED of the whole {\pmnt}
(restricting the sample to sources above the $4\,\sigma$ detection level at 143\,GHz). Therefore, while we cannot rule out the possibility of Galactic CO emission as the origin of the observed excesses at 100\,GHz for some of the sources of the catalogue, there is no evidence that this effect is relevant for the majority of sources in the {\pmnt}.
Finally, a third possibility is that the 100-GHz bump is a manifestation 
of the 
natural variation in
blazar SEDs at low and intermediate frequencies. Blazar spectra are the combination of emission from different components of the jets. Perhaps the excesses and deficits seen in Fig.~\ref{fig:HSS_massaro_fluxes} are simply a manifestation of these different components, and have no further relevance. It is also possible that the observed excesses at 100\,GHz in Fig.~\ref{fig:HSS_massaro_fluxes} are due to a combination of the three possibilities above.

\subsection{Statistical properties of the catalogues} \label{sec:statprop}

The main goals of this paper are to introduce the {\pmnt}, {\bpmnt}, and {\hpmnt} catalogues, to explain how they were obtained using \planck\ data, and to make them available to the community.
Table~\ref{tb:summary} summarizes some of the results of our investigations into cross-matching the {\pmnt} to external catalogues of radio sources, blazars, and Galactic cold cores, discussed in the previous sections for sources outside the \GAL070\ mask.\footnote{The total number of {\pmnt} sources outside the \GAL070\ Galactic mask is 18\,647.} The table 
gives an approximate idea of the number of sources detected by \planck\ not covered in lower frequency surveys,\footnote{As seen in the previous sections, this number is not 
equivalent to an assessment of the reliability of the catalogue. See Sect.~\ref{sec:relcomp} for a full discussion of reliability.} of the level of contamination by Galactic objects for the different \planck\ channels, and of the fraction of bright {\pmnt} sources previously identified as blazars.
A full statistical study of the catalogue is out of the scope of this work; however, to illustrate the potential of the catalogues we outline here their main statistical properties and the science that can be derived from them.

\begin{table*}[htbp!]
\newdimen\tblskip \tblskip=5pt
\caption[]{Brief summary of the statistical results of cross-matching the \pmnt to external catalogues of radio sources, blazars, and Galactic cold cores. Only sources detected above the $4\,\sigma$ level outside the \GAL070\ Galactic mask are
considered here.}
\label{tb:summary}
\vskip -3mm
\footnotesize
\setbox\tablebox=\vbox{
\newdimen\digitwidth
\setbox0=\hbox{\rm 0}
\digitwidth=\wd0
\catcode`*=\active
\def*{\kern\digitwidth}
\newdimen\signwidth
\setbox0=\hbox{+}
\signwidth=\wd0
\catcode`!=\active
\def!{\kern\signwidth}
\newdimen\pointwidth
\setbox0=\hbox{.}
\pointwidth=\wd0
\catcode`?=\active
\def?{\kern\pointwidth}
\halign{\hbox to 6.0cm{#\leaderfil}\tabskip 1em&
\hfil#\hfil& \hfil#\hfil& \hfil#\hfil& 
\hfil#\hfil& \hfil#\hfil& \hfil#\hfil& 
\hfil#\hfil& \hfil#\hfil& \hfil#\hfil\tabskip 0em\cr
\noalign{\doubleline}
\noalign{\vskip -1pt}
\omit\hfil Channel [GHz]\hfil& 30& 44& 70& 100& 143& 217& 353& 545& 857\cr
\noalign{\vskip 3pt\hrule\vskip 5pt}
Number of $\geq4\,\sigma$ detections& 1701& 1123& 1328& 2015& 2047& 1610& 943& 719& 798\cr
$4\,\sigma$ matched to CRATES (\%)& 70.4& 77.4& 80.0& 75.5& 68.9& 67.4& 57.3& 32.3& 22.4\cr
$4\,\sigma$ matched to GB6+PMN (\%)& 88.6& 90.8& 92.4& 91.3& 85.9& 83.8& 75.5& 60.9& 52.5\cr
$4\,\sigma$ matched to PGCC (\%)& 0.9& 0.8& 0.6& 1.6& 1.3& 7.3& 13.9& 18.8& 16.9\cr
$4\,\sigma$ matched to BZCAT5 (\%)& 54.0& 63.4& 64.0& 53.2& 50.0& 51.6& 47.1& 22.7& 11.8\cr
\noalign{\vskip 3pt\hrule\vskip 5pt}}}
\endPlancktablewide
\end{table*}

\subsubsection{Spectral indices and colour-colour plots} \label{sec:spindex}

\begin{figure*}[h]
   \centering
   \includegraphics[width=\textwidth]{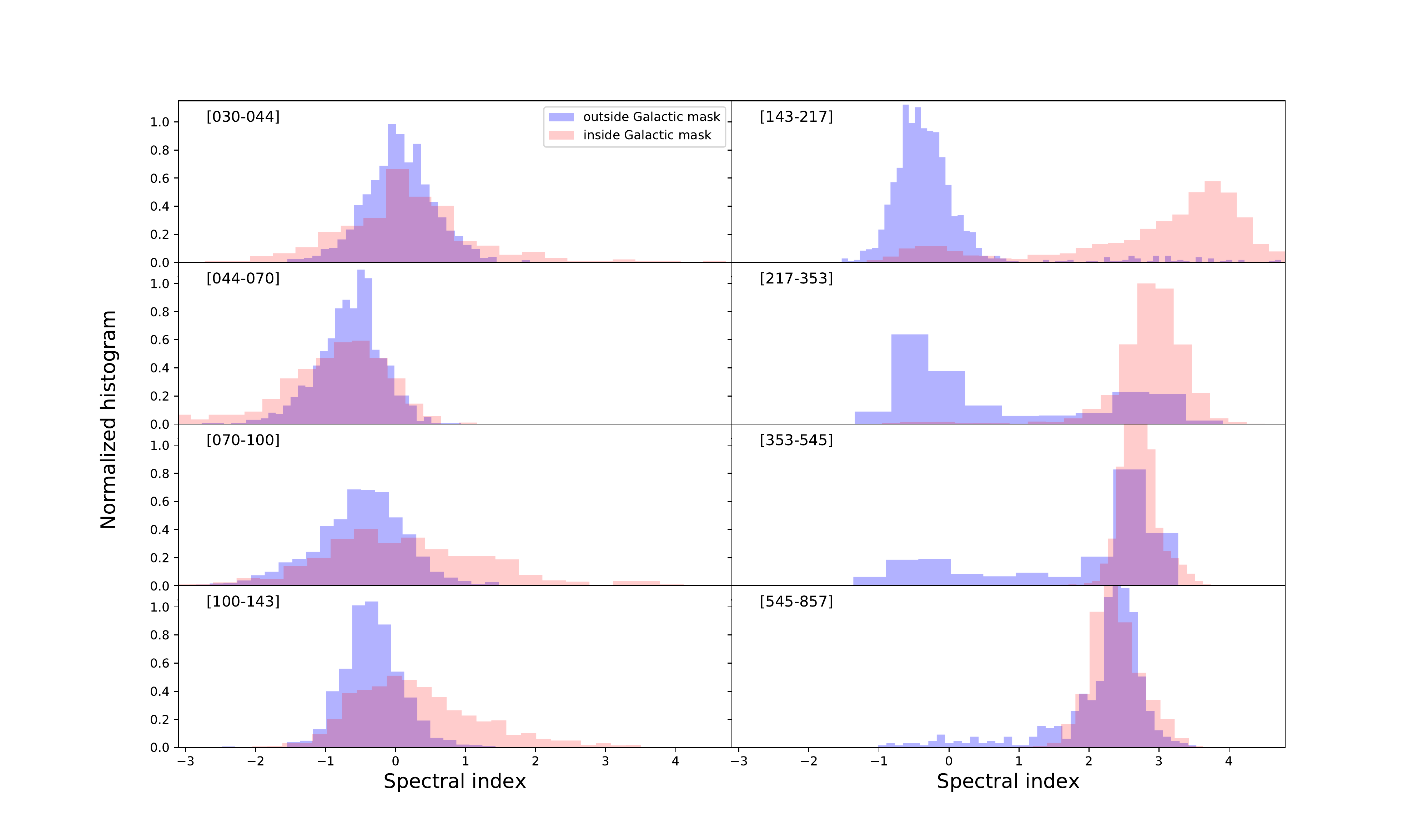}
   \caption{Normalized histograms of {\pmnt} spectral indices for different pairs of \planck\ frequencies.
   Only sources detected above the $4\,\sigma$ level in each pair of frequencies are considered.
   Sources within the \planck\ \GAL070\ mask are shown in red, whereas the sources outside the quoted
   Galactic mask appear in blue. Instead of the usual fixed-width bins, the histograms use \texttt{optBINS},
   an optimal adaptive data-based binning method that adjusts itself to optimally reflect the details of the structure
   of the underlying distribution
\citep{knuth}. }
              \label{fig:spec_indx}
    \end{figure*}

The distribution of spectral indices in the {\pmnt} is shown in Fig.~\ref{fig:spec_indx}. 
We have estimated the
spectral indices only for those sources detected above the $4\,\sigma$ level in each pair of frequencies. We distinguish
between (mostly) Galactic sources inside the \planck\ \GAL70\ Galactic mask and (probably) extragalactic sources outside the
mask. For the extragalactic sources, we again recover the high-frequency synchrotron steepening above 70\,GHz
anticipated by \cite{NEWPSstat} using WMAP-detected sources and confirmed in a series of papers \citep{planck2011-6.1,planck2012-VII,planck2013-p05,planck2014-a35,massardi11,massardi16}. As already found in the
analysis of the Planck Early Release Compact Source Catalogue \citep{planck2011-1.10} and confirmed later
with the PCCS and the PCCS2 \citep{planck2013-p05,planck2014-a35}, extragalactic sources with thermal emission
begin to appear only at 217\,GHz and above, a much higher frequency than expected before the launch of the
\planck\ mission. 

From these analyses, we can conclude that the {\pmnt} catalogue is an
excellent resource for studying the statistical properties of the non-thermal source population and the changes in
their emission properties with frequency at all the \planck\ channels (a frequency range poorly observed in the
past). The possibility to observe the emission of the same source across such a wide frequency range allows us to apply
additional selection criteria in order to identify important types of source. For example,
selecting sources with flat spectrum behaviour at relatively high frequency (143 and 217\,GHz) will mainly provide
us with a blazar-AGN subsample, as was seen in Sect.~\ref{sec:BZCAT}.

\begin{figure}[h]
  \centering
   \includegraphics[width=\columnwidth]{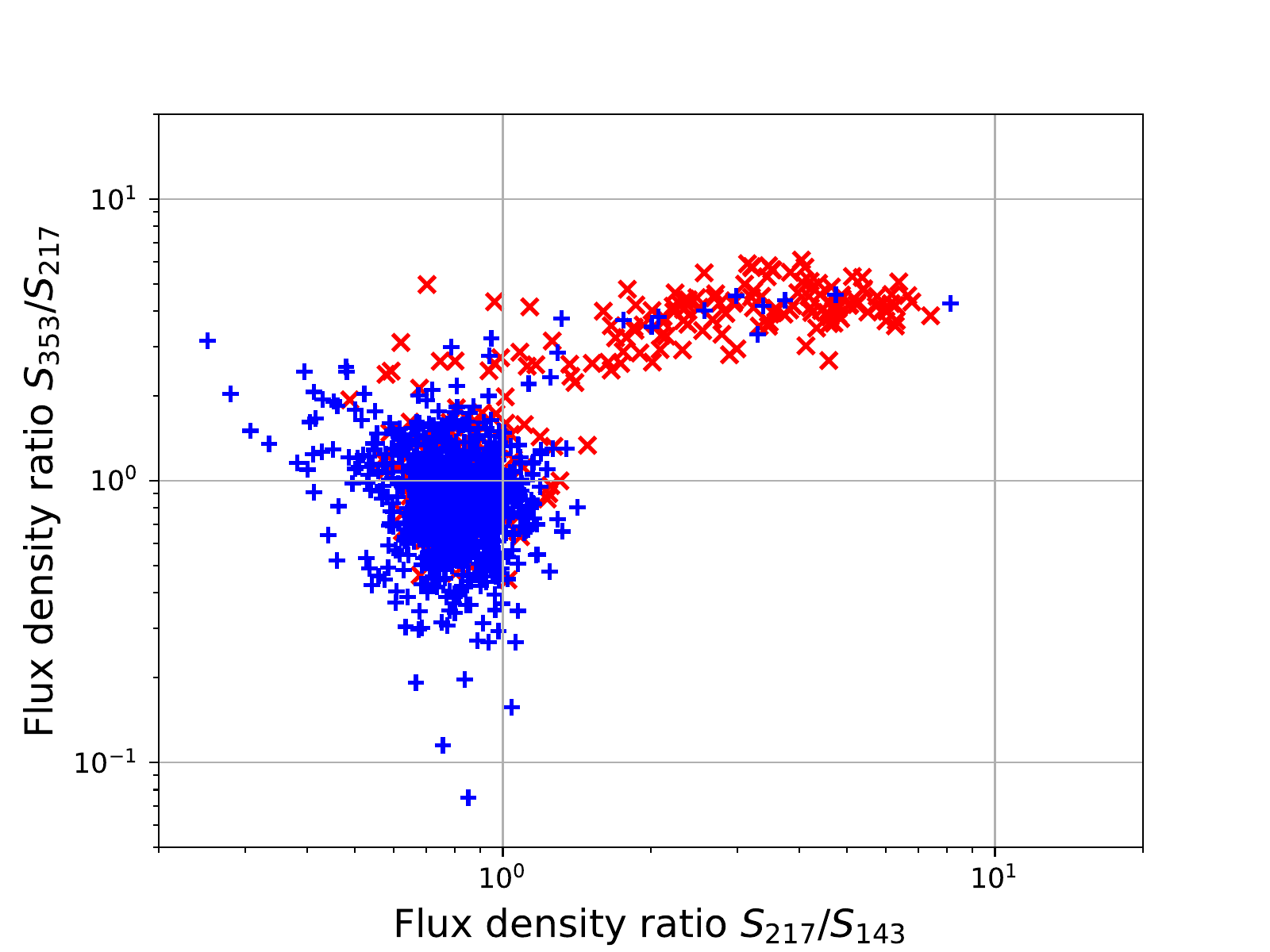}
   \caption{Colour-colour plot of the {\bpmnt} catalogue at a common frequency 217\,GHz.
The red crosses show the 278 {\bpmnt} sources inside the \planck\ \GAL070\ mask, while the blue pluses show the 1146 {\bpmnt} sources outside the same mask.
    We can see the non-thermal and thermal source populations of the {\bpmnt}; the non-thermal source population is dominant, as is expected from our selection criteria. }
              \label{fig:colour-colour}
\end{figure}

Fig.~\ref{fig:colour-colour} shows the colour-colour plot at the common frequency 217\,GHz for the sources in the {\bpmnt}. This figure is analogous to the top panel of figure~26 in \citep{planck2014-a35}.
The red crosses show the 278 {\bpmnt} sources inside the \planck\ \GAL070\ mask, whereas the blue pluses show the 1146 {\bpmnt} sources outside the same mask.
We can see the non-thermal and thermal source populations of the {\bpmnt}. Sources inside the
\planck\ \GAL070\ mask tend to have a thermal spectrum, whereas the sources outside the same mask are mainly non-thermal.
 The non-thermal population is much more numerous than the thermal population, thanks to our selection criteria at 30 and 143\,GHz.  This indicates that the {\bpmnt} is a good multi-frequency catalogue of bright non-thermal sources.

At frequencies higher than 353\,GHz, we can see a different behaviour between the whole {\pmnt} sample and the
{\hpmnt}. In the case of the {\pmnt}, most of the sources show clear thermal emission, which probably indicates
some kind of Galactic thermal contamination or the classical Eddington bias. However, this thermal emission 
dominates only in roughly half of the {\hpmnt}, and only for the 857-GHz channel. Taking into account that these
sources are visible in all the \planck\ channels, the best interpretation in this particular case is that they
are bright local galaxies, where we are able to observe both the radio and thermal emission.
Within a search radius of $5^{\prime}$ and outside the \GAL070\ mask, there are 23 identifications\footnote{Out of 75 {\hpmnt} sources outside the \GAL070\ mask.} of {\hpmnt} sources with IRAS compact sources \citep{IRAS,IRAS_GCAT}.\footnote{Among them are several well known local galaxies, such as Cen~A, M82, M87, and the Sculptor Galaxy.}

\subsubsection{Galactic and thermal sources in the catalogues} \label{sec:Galactic}

Fig.~\ref{fig:spec_indx} shows, using different colours, sources outside and inside the \planck\ \GAL070\ mask. This allows us to roughly distinguish between Galactic and extragalactic-dominated subsamples of the corresponding catalogues. In this section we briefly discuss the properties of the sources inside the \GAL070\ mask and quantify the impact of cold thermal sources that were unintentionally selected in our input sample.

The Galactic sources of the {\pmnt}, as seen in Fig.~\ref{fig:spec_indx}, behave similarly 
to the extragalactic ones below 143\,GHz, where sources emitting
thermally first appear. This early appearance 
of thermal emission
is probably due to the fact that most of the Galactic sources are much
brighter than the extragalactic ones.  Galactic sources that have a smooth transition between non-thermal- and
thermal-dominated channels are probably mostly planetary nebulae \citep[see][]{planck2014-XVIII} or supernova remnants \citep[see][]{planck2014-XXXI}. Furthermore, Galactic sources
showing a variation in their spectral behaviour around 100\,GHz can be potential candidates for studying CO line
emission. 
This suggests that the spectral index plot can be used as a
 stand-alone tool to distinguish Galactic sources in the catalogue. 
In addition, but less easily interpretable from the physical point of view, further selection criteria can be
devised based on a principal component analysis (PCA) of the SEDs (see for example Sect.~\ref{sec:PCA}).

As mentioned in Sect.~\ref{sec:cat0}, the use of the 143-GHz band as one of the selection channels for our input \planck\ sample carries the risk of including bright thermal sources in the {\pmnt}. 
In some cases, these are extragalactic sources that show both non-thermal synchrotron spectrum at low frequencies and thermal emission at high frequencies; in some other cases, they can be purely thermal sources that are bright enough to be detected above our selection threshold at 143\,GHz. 
In order to quantify the impact of thermal emission in the high-frequency channels of the {\pmnt}, we have
looked for sources detected at ${\rm S/N}\geq 4$ at 143\,GHz that have no counterpart in the CRATES catalogue \citep{CRATES} within a search radius of 32\parcm3 (the \planck\ FWHM at 30\,GHz). In order to focus on sources that are likely to be of
extragalactic origin, 
 we restrict our search to sources outside the \planck\ \GAL070\ Galactic mask and that are not in the 
 PGCC \citep{planck2014-a37}. There are 494 sources in the {\pmnt} that satisfy these criteria. Taking these, we use a very simple spectral index criterion in order to identify sources with potential thermal-like emission, namely those sources with spectral index between frequencies 143 and 217\,GHz $\alpha_{143}^{217} > 1$. 56 out of these 494 sources ($11.3\,\%$) have thermal spectra according to the $\alpha_{143}^{217} > 1$ criterion. This suggests that the degree of contamination of thermal sources in the whole {\pmnt} at high Galactic latitudes should be relatively small. However, the situation changes significantly if we focus on the brightest sources at 857\,GHz. If we repeat the analysis, but consider only sources with flux density above 1\,Jy at 857\,GHz, we find that almost half of the sources (47 out of 97 that have $S_{857} \geq 1\,$Jy and ${\rm S/N}_{143}\geq4$ at 143\,GHz, are outside the \GAL070\ mask and are not matched to CRATES) sources are thermal-like according to the $\alpha_{143}^{217} > 1$ criterion. This implies that the contamination from thermal sources is much more relevant in the bright source part of the 857 counts, and that it cannot be neglected even for regions outside the Galactic plane. 
 To a lesser extent, the same applies to the 545- and 353-GHz \planck\ channels. 
 Source number count plots (not included in this paper for the sake of brevity) at the high-frequency \planck\ channels ($\nu \geq 353$\,GHz) show a clear excess of sources above 1\,Jy with respect to the \cite{tucci11} model of radio source number counts.

\subsubsection{Principal component analysis} \label{sec:PCA}

Colour-colour diagrams
and fits to certain spectral laws (a modified blackbody, for example, or a power law with a given spectral index) are useful tools for source
classification. However, the complexity and arbitrariness of the classification rules can grow very quickly when the number of channels (or colours) is large. In that case, statistical procedures such as principal component analysis \citep[PCA, see for example][]{PCA} can help to identify trends and correlations, to reduce the dimensionality of the data and to ease the burden of automatic classification of sources in a catalogue.

PCA is mathematically defined as an orthogonal linear transformation that transforms the data to a new coordinate system, such that the greatest variance through some projection of the data comes to lie on the first coordinate (called the first principal component), the second greatest variance on the second coordinate, and so on. In our case, the data for each source are the standardized (that is, mean subtracted and normalized to unit variance) flux densities from 30 to 857\,GHz, and the intuitive meaning of the principal components is that of `generalized colours' that identify the main spectral trends. Operationally, finding the principal components is equivalent to solving the eigen-problem for the covariance matrix of the data (in our case, the covariance matrix is a $9 \times 9$ matrix computed by cross-correlating the 29\,400 spectral energy distributions we have in the {\pmnt}). If $\mathbf{X}$ 
is the $29\,400 \times 9$ matrix of standardized (normalized) flux densities of the catalogue, the principal components are given by
\begin{equation}
\mathbf{T} = \mathbf{X} \mathbf{W},
\end{equation}
\noindent
where $\mathbf{W}$ is a $9 \times 9$ matrix whose columns are the eigenvectors of $\mathbf{X}^{\sf T} \mathbf{X}$.
We perform PCA on the standardized (normalized) flux densities from 30 to 857\,GHz for all the 29\,400 targets in the input catalogue. In the standardization pre-processing step, the SED of each source is renormalized to unit variance.
This is because we are interested in the shapes of the SEDs, not in their individual amplitudes.  After this pre-processing standardization we calculate the covariance matrix of the samples and obtain the principal components through the eigenvectors of the covariance matrix. We find
that the two first components account for $86\,\%$ of the sample variance. Adding a third principal component we account for $91\,\%$ of the sample variance. This means that most of the spectral information contained in the nine frequencies of the catalogue can be coded in just two or three numbers. 

\begin{figure*}[h]
   \centering
   \includegraphics[width=\textwidth]{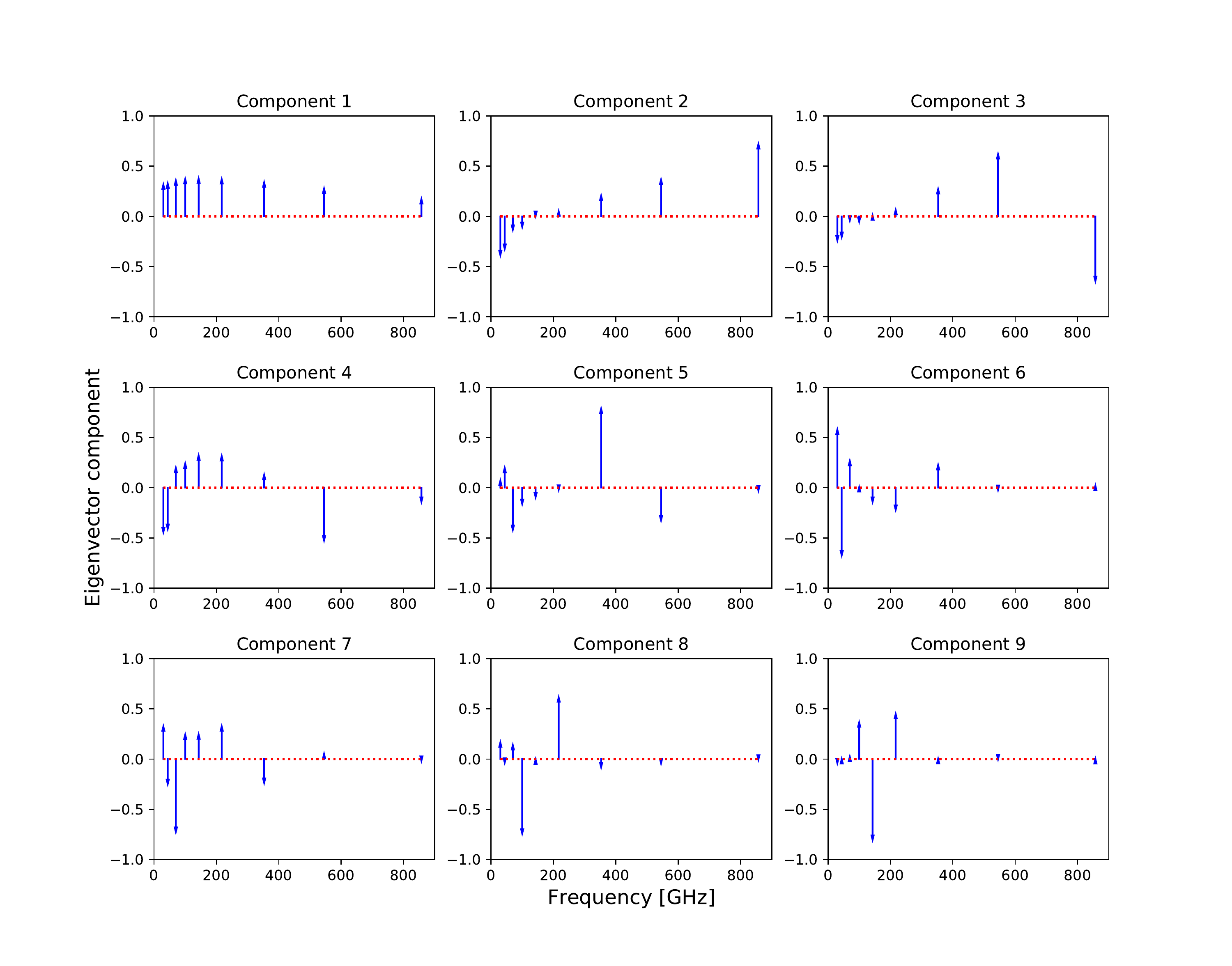}
   \caption{Components, as a function of frequency, of the nine loading vectors (eigenvectors) that map the normalized SEDs into the principal component space.}
              \label{fig:eigenvecs}
    \end{figure*}

It is not always easy to find an intuitive interpretation for these generalized colours. Fig.~\ref{fig:eigenvecs} shows the
components, as a function of frequency, of the nine eigenvectors calculated from the {\pmnt}. The two first components, which as we have just mentioned account for $86\,\%$ of the spectral information content, could be interpreted
as more or less a flat spectrum with some preponderance of radio emission (which was to be expected, considering our selection criterion at 30 and 143\,GHz), plus a component that peaks towards the high frequencies (probably thermal dust contamination). 
The third vector suggests some kind of bump around 545\,GHz, which may be associated with a Galactic cold dust component with $T\simeq10\,$K (or warmer sources at $z>0$). The rest of the components are much more difficult to interpret, but carry much less information about the spectral behaviour of the catalogue. 

\begin{figure}[h]
   \centering
   \includegraphics[width=\columnwidth]{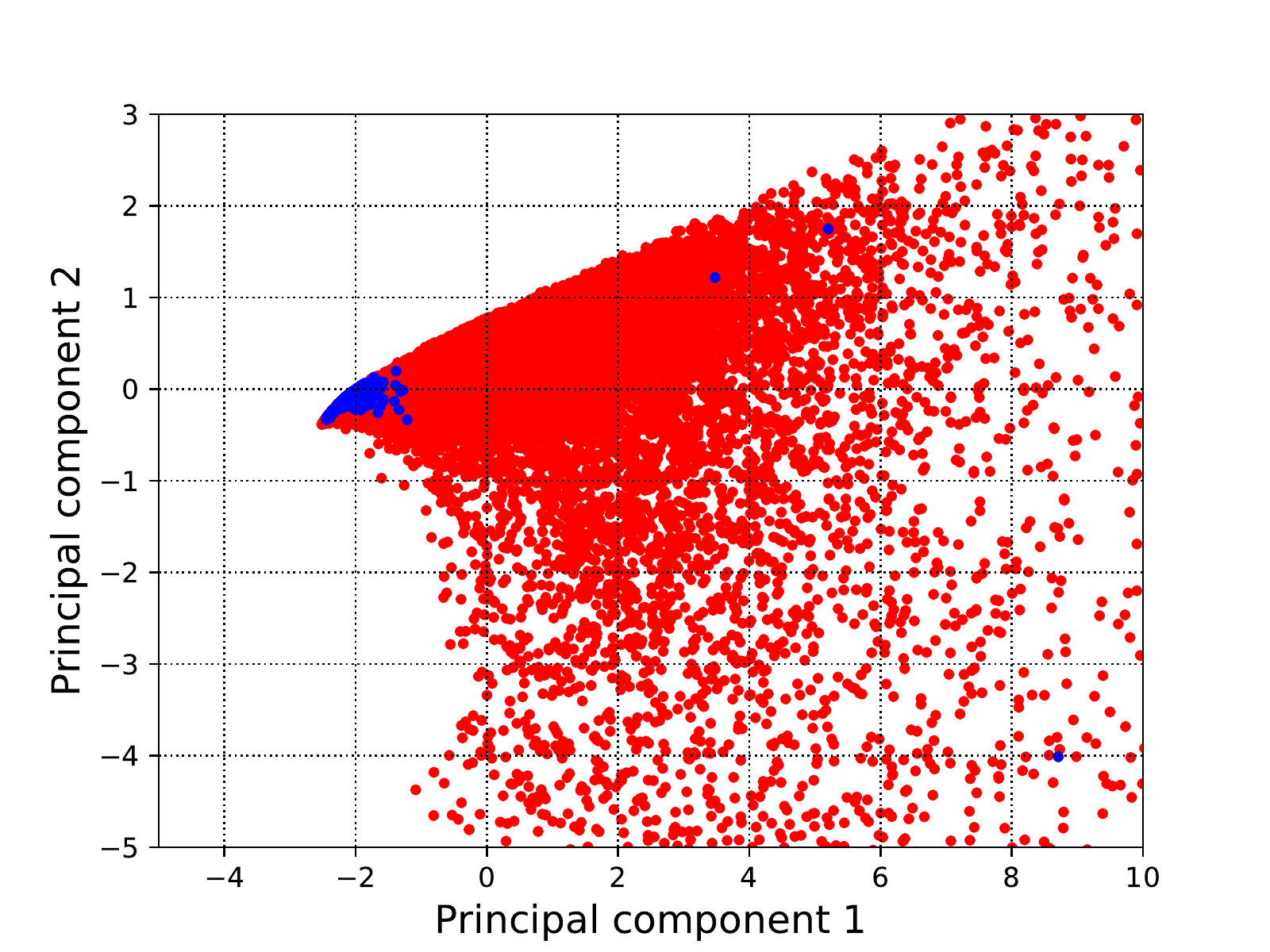}
   \caption{First two principal components for the whole input catalogue (red points) and a subset of sources selected for their strongly thermal spectra (blue dots).}
              \label{fig:PCA2D}
    \end{figure}

The utility of the PCA analysis can be seen in Fig.~\ref{fig:PCA2D}, which shows the distribution of the two first principal components for the whole catalogue (red dots). 
 We can now look for particular subsets of sources and see whether they lie in a restricted region of the diagram or not.

As an example, we have selected a group of bright Galactic sources by looking for coincidences between our catalogue and the PGCC \citep{planck2014-a37}. 
Galactic cold clumps are expected to have purely thermal spectra.
There are 3150 matches within a $5^{\prime}$ search radius between our catalogue and the PGCC. If we also restrict ourselves to bright sources that are detected with ${\rm S/N}>4$ at 353, 545, and 857\,GHz in our catalogue,
we keep 1882 objects out of 29\,400.
The sources selected in this way are shown as blue dots in Fig.~\ref{fig:PCA2D}, where they appear strongly clustered in a thin, elongated structure, centred around $(-2.32,0.14)$ in the principle component plane. This suggest that it should be possible to identify sources with strong thermal emission just by applying an automated classification algorithm in the space of the first few principal components. 

\begin{figure*}[h]
 \centering
\includegraphics[width=\textwidth]{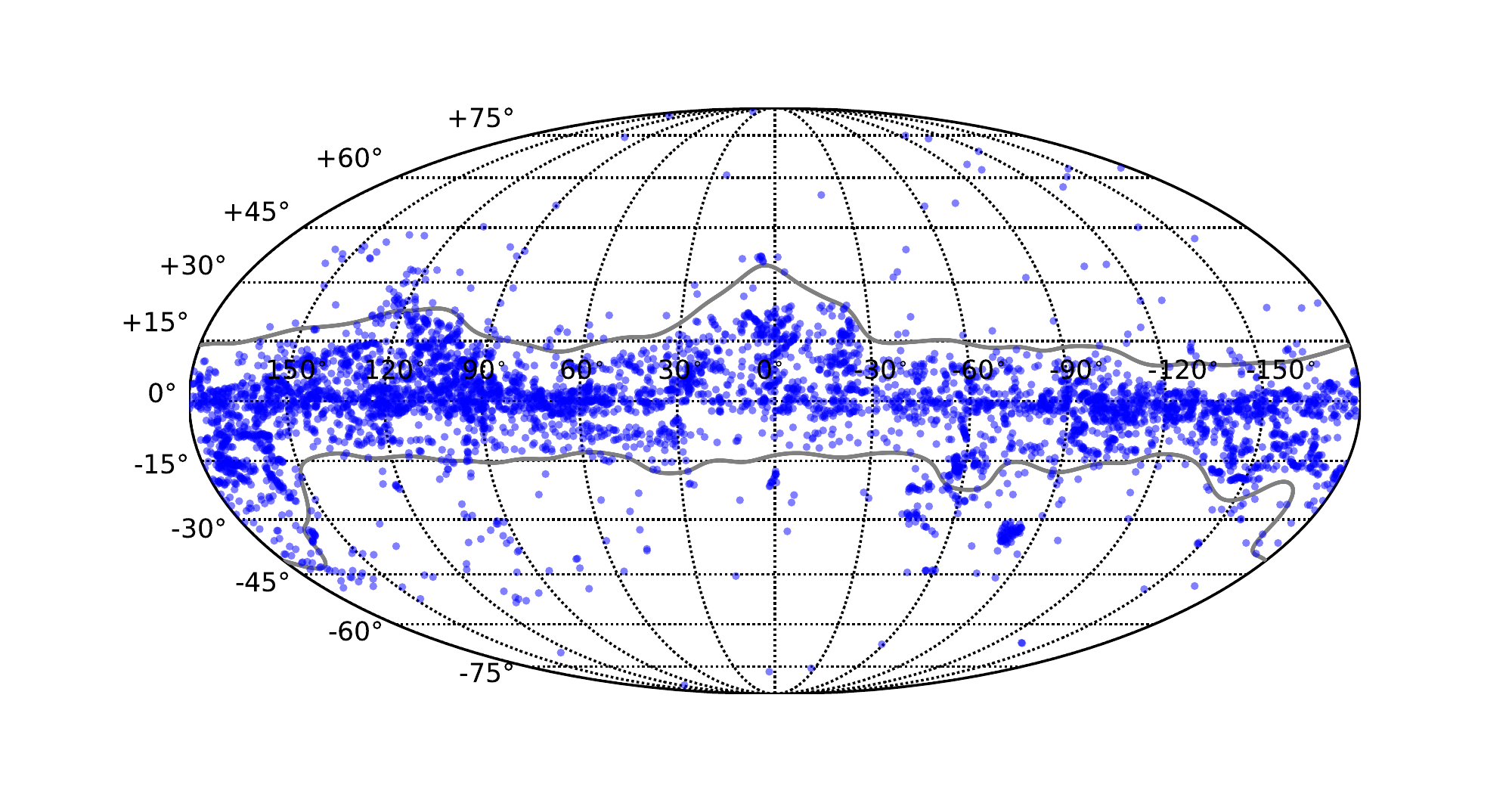}
\caption{Distribution on the sky of sources that were classified as thermal-like by the $k$-NN criterion on the first two principal components. The \planck\ $70\,\%$ Galactic mask \GAL070\ is superimposed in grey for comparison. The blob of sources around $l = -79\pdeg5$, $b=-32\pdeg9$ corresponds to the Large Magellanic Cloud.
}
\label{fig:knndist}
\end{figure*}

\begin{figure}[h]
 \centering
\includegraphics[width=\columnwidth]{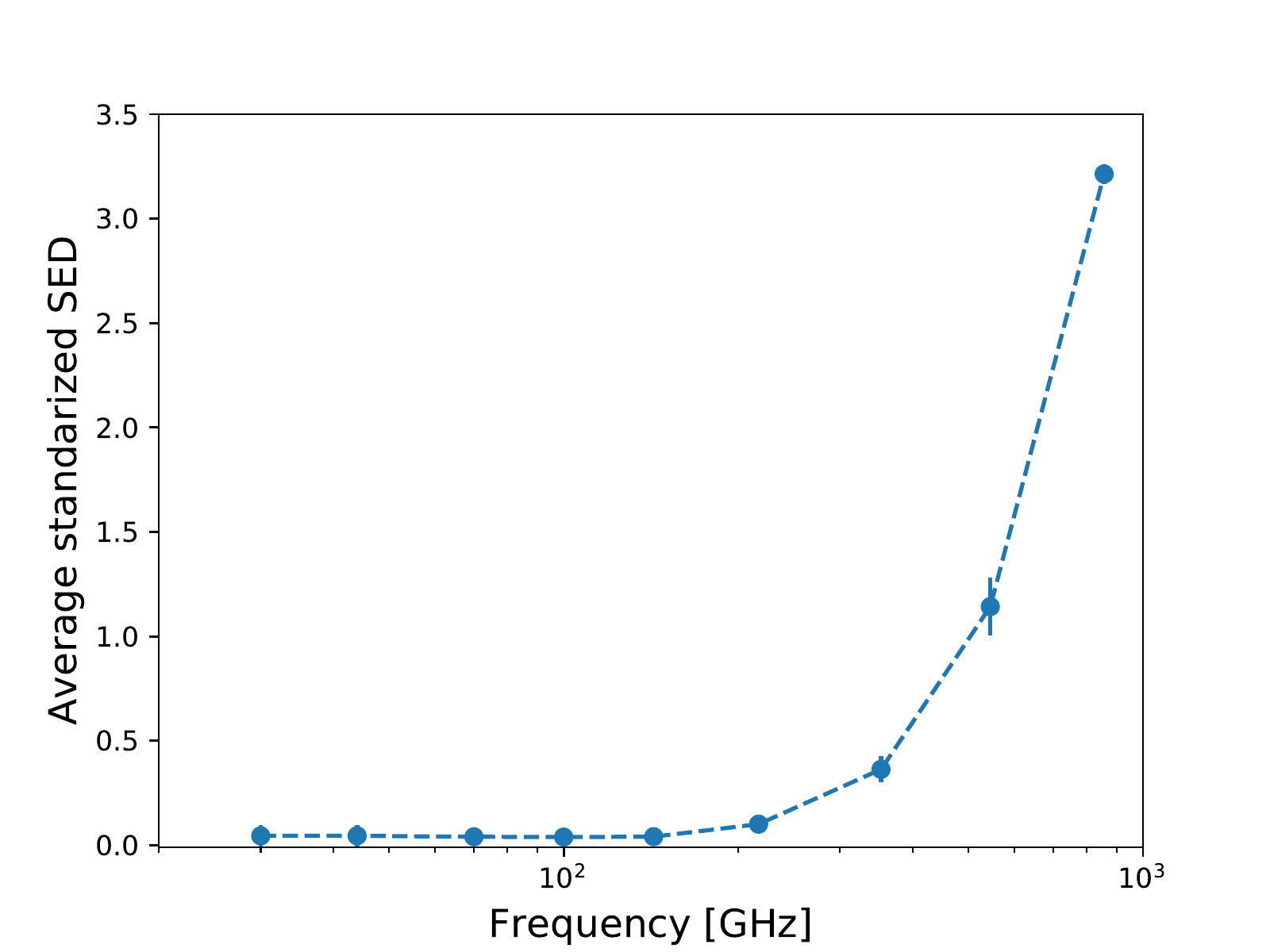}
\caption{Average standardized spectral energy distribution of the thermal-like sources selected by the $k$-NN criterion on the first two principal components. }
\label{fig:knnSED}
\end{figure}

The $k$-nearest neighbours algorithm \citep[$k$-NN,][]{knn1,knn2} is a non-parametric pattern recognition and classification method that provides an intuitive way to classify a $n$-dimensional sample of objects into two or more categories, starting from a training subsample such as the one we have described above. An object is classified by a majority vote of its neighbours, with the object being assigned to the class most common among its $k$ nearest neighbours ($k$ is a positive integer, typically small, and must be an odd number in order to avoid the possibility of having a tie in the vote). We have applied the $k$-NN method, with $k=5$, to the whole catalogue, restricting the analysis to only the two first principal components.\footnote{We have tried also other values, $k=3,7,9$ and the results do not change significantly.} According to the $k$-NN classificator, there are 4829 objects that should be classified in the same category of thermal-like sources. Fig.~\ref{fig:knndist} shows the distribution on the sky of these sources; for comparison, the \planck\ $70\,\%$ Galactic mask is outlined in grey. As we can see, the automatically selected sources lie around the Galactic plane and the Magellanic Clouds. Fig.~\ref{fig:knnSED} shows the average, standardized spectral energy distribution of the selected sources, where we can see an almost purely thermal spectrum with a minor contribution from synchrotron (or maybe free-free) emission in the lower frequencies. This result is in agreement with the physical interpretation we have given above for the two first principal components.

These basic examples should help to demonstrate the potential and the power of the PCA approach. Just by looking at two
principal components, instead of using a complicated set of colour-based rules (as in the standard approach), we
are able to identify -- and, thus, exclude -- sources showing a characteristic thermal dust spectrum. The
same approach could be also easily applied to the first three principal components instead of only the first two.
In general, this technique could be used to flag sources that have
extreme spectral behaviour before performing the statistical analysis of the catalogue of sources of greatest
interest for the purposes of each specific investigation (e.g. non-thermal ERSs, in this case).

\section{The catalogue: access, content and usage} \label{sec:contents}

\begin{figure*}[h]
 \centering
\includegraphics[width=\textwidth]{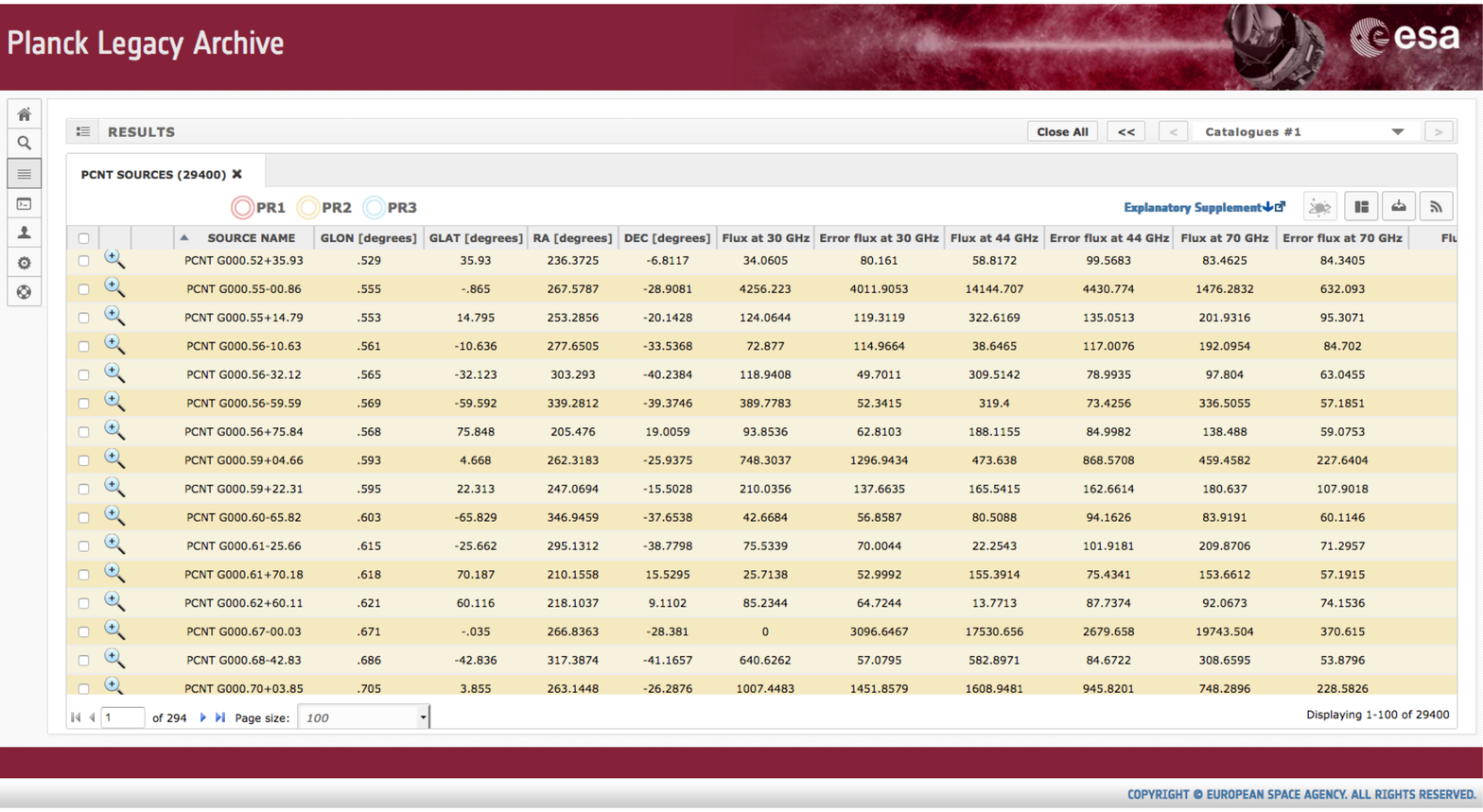}
\caption{Screenshot from the ESA  Planck Legacy Archive \pmnt\ portal.
}
\label{fig:screenshot}
\end{figure*}

The {\pmnt} catalogue 
is available from the Planck Legacy Archive.\footnote{\url{http://pla.esac.esa.int/pla}}
The name of the catalogue file in the Planck Legacy Archive is \texttt{COM\_PCCS\_PCNT\_R2.00.fits}.
 Fig.~\ref{fig:screenshot} shows a screenshot of the Planck Legacy Archive web interface to the \pmnt. 
 The {\pmnt}
contains the coordinates, flux densities, flux density errors, and MTXF S/N for 29\,400 sources. Here we summarize the catalogue contents.
\begin{itemize}
\item Source identification: \texttt{MAME} (e.g. {\pmnt}).
\item Position: \texttt{GLON} and \texttt{GLAT} contain the Galactic coordinates, and \texttt{RA} and \texttt{DEC} give the same information in equatorial coordinates (J2000).
\item Flux density: the estimates of flux density for the nine  \planck\ frequencies (\texttt{Err\_Flux\_xxx}), in mJy, and their associated uncertainties (\texttt{Err\_Flux\_xxx}). The string \texttt{xxx} contains the frequency value.
\item MTXF signal-to-noise ratio for the nine frequencies: \texttt{SNR\_xxxGHz\_MTXF}.
\end{itemize}
To facilitate the usage and scientific exploitation of the catalogue, the {\pmnt}
provides in addition seven different flag columns:
\begin{itemize}
\item \texttt{PCNTb}, indicating sources belonging to the {\bpmnt};
\item \texttt{PCNThs}, indicating sources belonging to the {\hpmnt};
\item \texttt{Group\_Flag}, indicating sources that are found inside a 30-GHz beam area as indicated in Sect.~\ref{sec:blend};
\item \texttt{Matched\_to\_PGCC}, indicating sources that are matched, within a 5\arcm\ radius, to PGCC sources;
\item \texttt{Matched\_to\_CRATES}, indicating sources that are matched, within a 
32\parcm3 radius, to CRATES sources;
\item \texttt{Matched\_to\_BZCAT5}, indicating sources that are matched, within a 7\arcm\ radius, to the BZCAT5 catalogue of blazars;
\item \texttt{from\_030\_input}, indicating sources that appear only at 30\,GHz in the input catalogue (see Sect.~\ref{sec:cat0} for further details).
\end{itemize} 

It should be noted that there is no column that contains the coordinate uncertainties for each source. The errors in position are 
inherited from the same Mexican hat wavelet technique used for the construction
of the PCCS2 and therefore can be computed using the same procedure as described in \cite{planck2014-a35}.

Additional information about the catalogue content and format can be found in the  FITS file headers. 

\section{Conclusions} \label{sec:conclusions}

The Planck Multi-frequency Catalogue of Non-thermal (i.e. synchrotron-dominated) Sources observed between
30 and 857\,GHz by the the ESA \planck\ satellite mission has been produced using the full mission data. This is the
first fully multi-frequency \planck\ catalogue of compact sources that covers all nine of the \planck\ frequency
bands.

We constructed our catalogue starting from a set of candidates that are blindly detected with Mexican hat filtering
on the \planck\ full-sky maps at 30 and 143\,GHz. From this input sample we selected those source
candidates with signal-to-noise ratio ${\rm S/N}>3$. The number of source candidates in this \planck\ input sample is
29\,400. We then ran a multi-frequency filtering (the Matrix Multi-filters) on the nine \planck\
frequency channels. The resulting all-sky \pmnt\ catalogue lists the positions, flux densities, and flux density uncertainties of 29\,400 compact
sources at all nine \planck\ frequencies. The {\pmnt} flux densities agree with previous \planck\
observations in the PCCS2 \citep{planck2014-a35}, but our
catalogue goes deeper in flux density limits from 30 to 143\,GHz (down to around 100--300\,mJy). Moreover, the
{\pmnt} is a band-filled multi-frequency catalogue over the whole \planck\ frequency range.  Therefore,
it is ideal for studying SEDs between 30 and 857\,GHz for thousands of compact
sources and investigating the statistics of their properties. An indication of the reliability of this multi-frequency
catalogue for statistical studies is given by the preliminary number counts calculated with sources in the {\pmnt}
and detected outside the Galactic \GAL070\ mask;
these are in very good agreement with the number counts predicted by the C2Ex cosmological
evolution model of ERSs described in \cite{tucci11}.

Although the selection criteria chosen for our \planck\ input catalogue tends to favour the selection of
non-thermal, i.e. synchrotron-dominated, sources, the {\pmnt} also selects many bright dusty compact sources,
the majority of them of Galactic origin. In order to provide a purer set of non-thermal sources, we also define a
subsample of the {\pmnt} whose components are detected with ${\rm S/N}>4$ at both 30 and 143\,GHz. This Bright Planck
Multi-frequency Catalogue of Non-thermal Sources ({\bpmnt}) contains 1424 sources, of which 1146 are found to
lie outside the \planck\ $70\,\%$ \GAL070\ Galactic mask.  Remarkably, 913 out of the 1424 sources of the
{\bpmnt} have counterparts within 7$^\prime$ (the \planck\ FWHM at 143\,GHz) in the BZCAT5 catalogue of blazars
\citep{massaro15}. 
Of these matches, 832 lie outside the \planck\ \GAL070\ Galactic mask.
Thus, the {\bpmnt} contains not only flat-spectrum ERSs (mainly classified as blazars), but also
a minority of Galactic sources, with spectra dominated by thermal dust emission.

Finally, we also flag the high-significance subsample ({\hpmnt}), a subset of 151 sources that are
detected with ${\rm S/N}> 4$ in all nine \planck\ channels. The {\hpmnt} contains high S/N SEDs
between 30 and 857\,GHz for 72 known blazars \citep{massaro15}, 60 of which lie outside the \planck\ \GAL070\
Galactic mask.

We release for the scientific community the full {\pmnt} and the bright subsample {\bpmnt} catalogues, and
also the {\hpmnt}, with the corresponding sources flagged in the {\bpmnt}. We also include in the catalogues nine
columns containing the results of a principal component analysis of the SEDs in our catalogue. We suggest that the
principal component analysis will prove useful for developing automated source classification criteria and,
as an example, we show that sources with a characteristic thermal spectrum, easily identified by using
only a simple classifier based on the two first principal components.

\begin{acknowledgements}
The Planck Collaboration acknowledges the support of: ESA; CNES, and CNRS/INSU-IN2P3-INP (France); ASI, CNR,
and INAF (Italy); NASA and DoE (USA); STFC and UKSA (UK); CSIC, MINECO, JA, and RES (Spain); Tekes, AoF, and CSC
(Finland); DLR and MPG (Germany); CSA (Canada); DTU Space (Denmark); SER/SSO (Switzerland); RCN (Norway); SFI
(Ireland); FCT/MCTES (Portugal); and ERC and PRACE (EU). A description of the Planck Collaboration and a list of
its members, indicating which technical or scientific activities they have been involved in, can be found at
\href{http://www.cosmos.esa.int/web/planck/planck-collaboration}{http://www.cosmos.esa.int/web/planck/planck-collaboration}. 
This research has made use of data from the OVRO
40-m monitoring program, which is supported in part by NASA grants NNX08AW31G, NNX11A043G, and
NNX14AQ89G and NSF grants AST-0808050 and AST-1109911. We also thank the Spanish MINECO for partial financial
support under project AYA2015-64508-P and funding from the European Union’s Horizon 2020 research and innovation
programme (COMPET-05-2015) under grant agreement number 687312 (RADIOFOREGROUNDS). This work was supported in part
by the ESAC Science Faculty award ESAC-362/2015. 
This research has made use of the NASA/IPAC Extragalactic Database (NED), which is operated by the Jet Propulsion Laboratory, California Institute of Technology, under contract with the National Aeronautics and Space Administration.
\end{acknowledgements}

\bibliographystyle{aat} 
\bibliography{pmtxf,planck_bib} 

\begin{appendix}
\section{Theory of the matched multi-filters} \label{sec:appendix}

\subsection{Multi-frequency detection}

Let us consider the problem of detecting and estimating the flux density of a source that is located at a given point of the sky defined by the position vector $\vec{b}$.\footnote{For the sake of clarity, throughout this paper we  use a notation according to which bold italic $\vec{x}$ symbols will denote vectors on the sphere (in either the spatial or Fourier domains), whereas plain boldface $\mathbf{f}$ symbols will denote vectors and matrices in the electromagnetic frequency domain.}  Let the source be measured over positions $\vec{x}$ at $N$ different frequency channels with \textit{different} angular beam responses $\lbrace \tau_k \left( \vec{x} \right) \rbrace = \left[ \tau_1 \left( \vec{x} \right), \ldots , \tau_N \left( \vec{x} \right) \right]$. For convenience, let us normalize the angular responses so that $\tau_k\left( \vec{0} \right) =1$, $k=1,\ldots,N$. Then the flux density of the source at a given position and channel can be written as
\begin{equation} \label{eq:model0}
S_k \left( \vec{x} \right) = A_k \tau_k \left( \vec{x}-\vec{b} \right),
\end{equation}
\noindent
where $A_k$ is the unknown amplitude of the source in the
$k$th channel. We assume no prior knowledge about the values of $A_k$, but we consider that the $\tau_k \left( \vec{x} \right)$ profiles are sufficiently well characterized by the instrument's optics and the
telescope's observing strategy. Now let the source be embedded in a mixture $n_k \left( \vec{x} \right)$ of CMB, Galactic and extragalactic foregrounds, and instrumental noise. We will refer to this mixture as the generalized noise.
 Let us model this generalized noise as a random process whose second-order statistics are either known a priori or can be directly estimated from the data:\footnote{If the sources are not very numerous or extremely bright, a good approximation of the power spectrum of the generalized noise is given just by the cross-power spectrum of the images. Note, however, that since the ingredients of the mixture of CMB and foregrounds depend on the position on the sky, the generalized noise must be computed locally rather than globally.}
\begin{equation} \label{eq:noise_ps}
\langle n_k \left( \vec{q} \right) n^{*}_l \left( \vec{q}^{~\prime}
\right) \rangle = P_{kl} \left( \vec{q} \right) \delta^2 \left( \vec{q} -
\vec{q}^{~\prime} \right),
\end{equation}
where $\mathbf{P} = (P_{kl})$ is the cross-power spectrum matrix of the generalized noise, $\vec{q}$ is a vector in Fourier space\footnote{It is 
convenient, both theoretically and computationally, to define the filters in Fourier space, and so
the $\vec{q}$ vector can be interpreted as a wavevector for Fourier analysis. We do not use the standard 
$\vec{k}$ notation in order to avoid confusion with the $k$ index in Eq.~(\ref{eq:model0})
and later. 
Since we are working with flat-sky patches, we can use with the planar Fourier transform instead of spherical harmonics. However, this is not essential for the discussion here, and all the formulae in this paper can be easily adapted to the spherical domain.}
 and
the asterisk denotes complex conjugation. When the generalized noise is added to the source signal, the observed data becomes
\begin{equation} \label{eq:datamodel}
  D_k \left( \vec{x} \right) = S_k \left( \vec{x} \right) +
  n_k \left( \vec{x} \right).
\end{equation} 
\noindent
for $k=1,\ldots,N$.
A useful quantity to consider is the peak S/N of the source in channel $k$, defined as
\begin{equation}
\mathrm{\cal S}_k = \frac{A_k}{\sigma_k},
\end{equation}
\noindent
where $\sigma_k$ is the rms of the generalized noise in the $k$th channel.
A typical way to enhance the detectability of a source in an individual channel is to spatially filter the data with a suitable filter $\psi_k$, so that the rms of the filtered generalized noise, $\sigma_{w_k}$, is conveniently reduced and
the
peak signal-to-noise ratio of the source is maximized \citep[e.g.][and references therein]{Barreiro2003}. The \textit{gain} of the filter is defined as the ratio
\begin{equation}
g_{\psi_k} = \frac{A_{w_k} / \sigma_{w_k}}{{A_k}/{\sigma_k}} = \frac{A_{w_k} \sigma_k}{A_k \sigma_{w_k}}.
\end{equation}
In this equation, $A_{w_k}$ is the amplitude of the source in the $k$th channel after filtering. It is always possible to normalize the filter in such a way that the amplitude of the sources is not altered, and then the gain reduces to
\begin{equation}
g_{\psi_k} = \frac{ \sigma_k}{ \sigma_{w_k}}.
\end{equation}

The optimal linear filter regarding the gain, for an individual channel, is the well known \textit{matched filter} \citep[see for example][]{Turin}. The members of the Mexican hat wavelet family achieve in practice almost as good results as matched filters and have been thoroughly used in CMB analysis, for example in the construction of the PCCS2 \citep{MHWF,lopezcaniego07,lopezcaniego09,herranz10,planck2011-1.10,planck2013-p05,planck2014-a35}. More recently, the biparametric adaptive filter \citep[BAF,][]{BAF} was proposed, which reaches virtually the same gains as the matched filter, while allowing a simpler and faster implementation.

Multi-frequency detection aims to further improve the gain by taking advantage either of the distinctive frequency dependence of the individual sources, or the statistical properties of the generalized noise, or both. But knowledge about the frequency dependence of most sources is scarce.
For synchrotron-dominated SEDs it is typical to approximate the spectral behaviour  by a power law
\begin{equation} \label{eq:powlaw}
A_{\nu} \propto \left( \frac{\nu}{\nu_0} \right)^{-\alpha},
\end{equation}
\noindent
where $\nu_0$ is some fiducial frequency of reference and $\alpha$ is the spectral index of the source. Using this kind of parametrization, a multi-frequency detection technique was applied to WMAP data in \cite{lanz13}. However, the parametrization (\ref{eq:powlaw}) is valid only for certain frequency and flux density intervals.
 Even if most of the radio sources exhibit synchrotron power-law SEDs in the low-frequency regime, variations from one particular source to another are large enough to make it impossible to describe all of them with a single frequency law. Moreover, these sources show a rich complexity of behaviour, at intermediate and higher frequencies. Therefore we choose to use a multi-frequency detection method that does not make any a priori assumption about the frequency dependence of the sources, relying only on the multi-frequency statistical properties of the generalized noise to maximize the probability of detection of \emph{any} kind of point source embedded in it.

\subsection{Matched multi-filters} \label{sec:MTXF}

The matched matrix filters (MTXFs) approach was introduced in \cite{MTXFa} and further explored in \cite{MTXFb} as a solution to the problem of multi-frequency detection of extragalactic point sources when the frequency dependence of the sources is not known a priori. 
The MTXFs have already been used inside the Planck Collaboration to validate the LFI part of the PCCS catalogue \citep{planck2013-p05} and we will use them in this paper.  So let us
summarize the main aspects of MTXF theory.

Let $\Psi_{kl} (\vec{x})$ be a set of $N \times N$ linear filters, and for the data  images $D_k$
let us define the filtered images
\begin{eqnarray} \label{eq:filtered_field}
  w_k (\vec{x}) & = & \sum_l \int d\vec{x}^{\prime} \Psi_{kl} \left( \vec{x} -
  \vec{x}^{\prime} \right) D_l \left( \vec{x}^{\prime} \right)
  \nonumber \\
  & = & \sum_l
  \int d \vec{q} \, \, e^{-i \vec{q}  \vec{x}} \, \Psi_{kl} \left(
  \vec{q} \right) D_l \left( \vec{q} \right).
\end{eqnarray}
The quantity $w_k$ in equation (\ref{eq:filtered_field}) is,
therefore, the sum of a set of linear filterings of the data
$D_l$ described in (\ref{eq:datamodel}) by the filters $\Psi_{kl} (\vec{x})$.
The rms of the filtered data can be calculated as the square root of the variance
\begin{equation} \label{eq:variance}
  \sigma^2_{w_k} = \sum_{l} \sum_{m} \int d\vec{q} \, \Psi_{kl}
  \left(\vec{q}\right) \Psi^*_{km} P_{lm} \left(\vec{q}\right).
\end{equation}
In \cite{MTXFa} it was shown that
the set of filters $\mathbf{\Psi}$ that minimize the variance $\sigma_{w_k}$ for all
$k$,
without changing the values of the individual amplitudes $A_k$ of the sources (independently of what kind of frequency dependence they have)
is given by the matrix equation
\begin{equation} \label{eq:matrix_filters_eqs}
\mathbf{\Psi}^* = \mathbf{F} \mathbf{P}^{-1},
\end{equation}
\noindent
where
\begin{align}
   \mathbf{\Psi} & = (\Psi_{kl}),\,\, \mathbf{F} = (F_{kl}),\,\,\mathbf{P} = (P_{kl}), \nonumber \\
  \pmb{\lambda} & = (\lambda_{kl}),\,\,\,\mathbf{H} = (H_{kl}),
\end{align}
\noindent
are $N \times N$ matrices for any value of $\vec{q}$ and where
\begin{align} \label{eq:matrix_filters_eqs2}
  F_{kl} & = \lambda_{kl} \tau_l, \nonumber\\
  \pmb{\lambda} & = \mathbf{H}^{-1}, \nonumber \\
  H_{kl} & = \int d\vec{q} \, \, \tau_k \left(\vec{q}\right) P^{-1}_{kl}
  \tau_l \left(\vec{q}\right) .
\end{align}
This set of filters assumes a structure that
is best expressed in the form of a matrix equation, hence the
name matched matrix filters. For $N$ different channels, the matrix equation gives a set of $N \times N$ linear filters that operate over the channels to produce $N$ filtered images, one for each channel, where the gain factor of point sources is maximized.

It is straightforward to show that when the filters (\ref{eq:matrix_filters_eqs}) are applied to a set of $N$ images, at the position of a given compact source the filtered images take values $\left [w_1,\ldots,w_N \right]$, with  
\begin{equation} \label{eq:condition1b}
w_k =  \sum_l \int d\vec{q} \, \Psi_{kl} \left(\vec{q}\right) A_k \tau_l
  \left(\vec{q}\right) = A_k,
\end{equation}
\noindent
that is, the filters are \emph{unbiased} estimators of the flux density of the source in every one of the $N$ channels. 
This is guaranteed by the algebraic 
 structure of the filters, that  is such that the
 following orthonormality condition is satisfied:
\begin{equation} \label{eq:orthonormal}
  \int d\vec{q} \, \Psi_{kl} \left(\vec{q}\right) \tau_l
  \left(\vec{q}\right) = \delta_{kl}.
\end{equation}
This relation guarantees that the flux density of a point source in the $l$th channel does not leak into the $k$th filtered image. On the other hand, if there is some degree of statistical correlation between the generalized noise in the $l$th and $k$th channels,
the $H_{kl}$ terms in Eq.~(\ref{eq:matrix_filters_eqs2}) contribute to decreasing the filtered rms $\sigma_{w_k}$ to a level that is lower than the one achieved by single-frequency matched filters. As shown in \cite{MTXFa} and \citet{MTXFb},
it can be shown that for the filters (\ref{eq:matrix_filters_eqs}) the variance (\ref{eq:variance}) is minimized, i.e. the MTXFs are not only unbiased but also achieve statistical \emph{maximum efficiency} among the class of linear filters when the SED of the sources is not known a priori. 
In other words, the matrix filters remove noise not only by filtering in the scale domain, but also by cleaning out medium- and large-scale structures that are present in neighbouring channels, while leaving unaltered the flux densities of sources.
In the particular case where 
\begin{equation} \label{eq:uncorrelation}
P_{kl} = \delta_{kl} P_k, 
\end{equation}
\noindent
i.e.
where the noise is totally uncorrelated among channels, it can be seen
that the elements of the matrix of filters default to
\begin{equation} \label{eq:uncorrelation2}
  \Psi^*_{kl} \left(\vec{q}\right) 
  = \delta_{kl} \frac{\tau_k \left(\vec{q}\right) / P_k
    \left(\vec{q}\right)}{
    \int d\vec{q} \, \tau^2_k \left(\vec{q}\right) / P_k
  \left(\vec{q}\right)}.
\end{equation}
In that case the matrix of filters becomes a diagonal matrix whose
non-zero elements are the complex conjugates of the matched filters
that correspond to each channel. In the case of circularly symmetric
source profiles and statistically homogeneous and isotropic noise, the
filters are real-valued and the whole process is equivalent to filtering
each channel independently with the corresponding
single-frequency matched filter. 

\section{Practical considerations of the implementation of the matched multi-filters for the \pmnt } \label{sec:practical}

In \cite{MTXFb} the MTXFs were applied to realistic pre-launch simulations of the \Planck-LFI + 100-GHz channels. The work served as a testbed for the capabilities of the MTXF approach and provided several practical lessons about the implementation of the technique.

The first of these lessons can be derived analytically from Eqs.~(\ref{eq:matrix_filters_eqs}) to (\ref{eq:orthonormal}). When the generalized noise is uncorrelated among channels the matrix of filters defaults to a diagonal matrix whose
non-zero elements are the complex conjugates of the matched filters
that correspond to each channel. In the case of circularly symmetric
source profiles and statistically homogeneous and isotropic noise, the
filters are real-valued and the whole process is equivalent to filtering
each channel independently with the appropiate matched filter.

The second lesson learnt from \cite{MTXFb} is that when the MTXF are applied to a set of $N$ frequency channels ordered by increasing (or decreasing) frequency, $\nu_1 < \nu_2 < \ldots < \nu_N$, the gains tend to be higher for the intermediate channels, and smaller for the peripheral channels. For the channels that are at the limits of the interval ($k=1$ and $k=N$), the gain tends to reach values similar to those obtained by single frequency matched filters.  There is no obvious theoretical reason for this to happen, but intuitively it seems logical that the noise reduction works better when there is information
about the correlated noise at frequencies on either side (which of course was the same driving reason for \Planck\ to have frequency coverage on both sides of the CMB channels).

The third lesson is related to equations (\ref{eq:uncorrelation}) and (\ref{eq:uncorrelation2}). For the MTXFs to improve the gain factors with respect to the single-frequency matched filter, it is necessary to have some degree of correlation between the generalized noise in different channels. Correlation between generalized noise appears due to the presence of common physical processes on the sky: for example, synchrotron emission at 30\,GHz correlates with synchrotron emission at 44\,GHz.
This implies that it makes little sense to mix together in the analysis frequency channels that are too far apart in the electromagnetic spectrum. The matrix $\mathbf{P}$ tends to have a band structure whose elements decrease rapidly away from the diagonal. If we add too many channels to the MTXF we gain little in terms of noise reduction, but we gain in computational complexity and potentially harmful numerical effects. It is convenient to find a compromise for the number of channels to consider simultaneously: too few channels means we lose gain power, but too many channels can lead to computational and numerical problems. We have found that $N=4$ channels is a good compromise for the case of \planck.

A fourth practical consideration pertains to the implicit and explicit assumptions of the filtering technique. The two basic statistical assumptions at the core of the MTXF technique are: (a) that the generalized noise (CMB + diffuse Galactic and extragalactic foregrounds + instrumental noise) is described by a random process sufficiently well described by its second-order statistics as in Eq.~(\ref{eq:noise_ps}); and (b) that the sources are point-like and the beam point spread functions at all the observing frequencies are well known, so that the orthonormality condition (Eq.~\ref{eq:orthonormal}) is satisfied. The first assumption is typical of most detection/estimation methods used in CMB astronomy and it is usually met at least in regions of the sky where Galactic emission is not too strong. In the following we will assume that the generalized noise is sufficiently well-behaved, but it should be noted as a caveat that the performance of the filters, particularly regarding the accuracy of their photometric estimation, can degrade in the vicinity of bright regions, especially in and around the Galactic plane.

Regarding the characterization of the sources as point-like compact objects, problems can arise in two cases: (a) when sources are extended; and (b) when the angular beam responses $\tau_k (\mathbf{x})$ are not well characterized. In both cases the orthonormality condition (Eq.~\ref{eq:orthonormal}) is not met and signal can leak from one frequency channel to others, thus affecting the quality of the photometry of individual sources.\footnote{
The orthonormality condition is crucial here. The MTXF method is combining $N$ images, mixing them and producing $N$ new filtered images, each one corresponding to one of the original channels. For each of the output channels we wish to detect and to estimate the flux density of the sources \emph{corresponding to that channel, and only that channel} We do not wish, for example, to look at a 70-GHz source and get its flux density mixed with (for example) its flux density at 143\,GHz. The way to guarantee that this is not happening is to satisfy Eq.~(\ref{eq:orthonormal}): if this condition is met, we can be sure that the MTXF procedure is cleaning contamination without mixing the flux densities of the sources with their flux densities at neighbouring frequencies.}  Let us for the moment focus on the second case. \planck\ beams are reasonably well approximated by circularly symmetric two-dimensional Gaussians with effective beam widths, as described in \cite{planck2014-a01},
\cite{planck2014-a05} and \cite{planck2014-a08}, except for one frequency, the LFI 44-GHz channel.  The LFI 44-GHz optical layout is composed of three pairs of beams (\texttt{LFI24} to \texttt{LFI26}), two of them with effective ${\rm FWHM}\simeq30\arcm$ and a third with effective ${\rm FWHM}\simeq23\arcm$ \citep[see][]{planck2014-a05}. Moreover, the individual ellipticities of these beams do not compensate each other perfectly, leading to a non-circular total beam. The beam heterogeneity of the 44-GHz channel makes it difficult to combine it with other channels in the MTXF scheme. For this reason, in this paper the 44-GHz channel is treated separately from the other channels, as described below.

Taking these considerations into account, for this work we have run the MTXF on sets of four channels (30--100\,GHz, 44--143\,GHz, 70--217\,GHz, 100--353\,GHz, 143--545\,GHz, and 217--857\,GHz). Except for the obvious cases of 30 and 857\,GHz, we accept flux densities for the final catalogue only for frequencies that are intermediate.
That is, the 100-GHz flux density could only come from the 44--143-GHz or 70--217-GHz runs.
Every time a channel is intermediate to two or more runs of the MTXF (for example, 143\,GHz is intermediate for the runs 70--217\,GHz and 100--353\,GHz), we choose for the final catalogue the case with a higher average S/N. Exception is made for the 44-GHz channel, whose non-circular, non-Gaussian effective beam puts into question the orthonormality condition (Eq.~\ref{eq:orthonormal}). For the 44-GHz channel we have thus opted to adopt the following rule.
\begin{itemize}
\item The photometry at 44\,GHz is obtained from the 30--100-GHz run of the MTXF, as described above. Due to the non-ideality of the 44-GHz beam the flux density estimates for this channel will be less accurate than for the rest of the \planck\ mission. In particular, for this case we observe that the 44-GHz MTFX flux densities are underestimated with respect to previous \planck\ catalogues. 
\item The 44-GHz MTXF flux densities must be corrected for this effect. We have 
used the PCCS2 sources as calibrators to obtain
an average correction (only for the 44-GHz channel).
The PCCS2 flux densities have been extensively checked and calibrated by the \planck\ team \citep{planck2014-a35} and thus we trust PCCS2 flux densities to be reliable calibrators, particularly taking into account that we are using the same sky images.
 The correction factor is obtained by calculating the average ratio between the 44-GHz PCCS2 and the 44\,GHz MTXF flux densities for sources with 44-GHz PCCS2 flux densities $> 500\,$mJy. The resulting factor is 1.36.
 We use this average ratio as a multiplicative calibration factor to obtain the final 44-GHz MTXF flux densities. 
\item In order to avoid contamination from the non-Gaussian sources at 44\,GHz leaking into the neighbour channels at 30 and 70\,GHz, we have performed a separate run of the filters using only the 30-, 70-, 100-, and 143-GHz bands. Flux densities at 30 and 70\,GHz come from this particular run.
\end{itemize}
For the rest of the channels (from 100 to 857\,GHz) we follow the more general rule described above.

\subsection{MTXF photometry} \label{sec:photometry}

By construction, MTXFs are normalized in such a way that the value of the filtered images at the position of a given source in a given frequency channel $k$ is an unbiased estimator of the source amplitude in that channel, $A_k$, as defined in Eq.~(\ref{eq:model0}), and in the same physical units as the input maps. Therefore the MTXF photometry, equivalent to the \texttt{DETFLUX} column in the PCCS and the PCCS2 \citep[see a description of the various methods used for \planck\ photometry in][]{planck2013-p05,planck2014-a35}, is obtained by just taking the value of the central pixel of the source and transforming it to flux density units (namely mJy).

While the extraction
of the flux densities is straightforward,
the true positions of the sources are not perfectly known. Since the input catalogue is constructed from a blind search on maps that are affected by foreground contamination and instrumental noise, there will be an unavoidable positional uncertainty, especially for the faintest sources. In order to deal with this uncertainty, we give as an estimation of $A_k$ the value of the local maximum of the filtered image inside a circle of radius equal to the FWHM at frequency $k$ centred at the coordinates of each target in the input catalogue.  This gives us a better chance of catching the true position of the sources; however, on the other hand it makes MTXF photometry more sensitive to 
positive fluctuations of the background, which could lead to overestimates in the photometry. We will discuss this problem in more detail in Sect.~\ref{sec:fluxval}.

As an estimate of the photometric error, we provide the rms of the filtered generalized noise, $\sigma_{w_k}$, calculated in a circular ring with inner radius of twice the FWHM (for frequency $k$) and outer radius of 5 times the FWHM, centred at the coordinates of the target.

Other photometric estimators, such as aperture photometry, can be useful for certain purposes, but cannot be directly applied to images filtered with the MTXF. This is because the output MTXF filtered images are zero-mean and the profiles of the filtered sources depend non-trivially on the statistics of the local foregrounds, which vary strongly across the sky. Additionally, application of any further filter (including aperture photometry) to optimally filtered maps results in a degradation of the S/N.  Aperture photometry must therefore be calculated on the non-filtered images, where the S/N enhancement of the MTXF has not been gained. For this reason and although aperture photometry can be more robust than MTXF photometry for the case of extended sources, we do not provide \texttt{APERFLUX} values for this catalogue.

\subsection{Blending effects} \label{sec:blend}

Since the {\pic} is selected at two channels (30 and 143\,GHz), it is possible that two or more targets selected at the channel 
with better angular resolution (143\,GHz in this case) lie inside the same 30-GHz beam.
In order to quantify the rate of occurrence of this kind of coincidence in our catalogue, we have performed an internal matching of the positions of the {\pic} using a search radius of 32\parcm293 \citep[the \planck\ Reduced Instrument Model FWHM at 30\,GHz;][]{planck2014-ES}. 
We find 5130 groups (affecting a total of 9125 sources with ${\rm S/N}\geq1$ at 30\,GHz) within that search radius, most of them containing two sources, but some of them with as many as 35 closely clustered positions. The extreme cases are located around the Galactic plane and the Magellanic Clouds. 
To a lesser extent, the same effect applies to the intermediate 44-, 70-, and 100-GHz channels. When this occurs, the \planck\ multi-frequency catalogue of non-thermal sources will contain different entries for low-frequency sources that would be otherwise identified as single objects. In principle, each one of these
entries will show a flux density that could be contaminated by overlap with the other targets in the vicinity convolved by the \planck\ beam at the corresponding observation frequency. Fortunately, we obtain the photometry of {\pmnt} sources after filtering with the MTXF method. By construction, MTXFs tend to have a deconvolving effect on point sources at lower frequencies, at least in regard to photometric estimations. 

In order to check the effect of source blending on MTXF photometry 
we have performed 1000 bootstrap simulations, injecting sources in 
regions devoid of bright sources in
the \planck\ maps. The distribution of flux densities of the simulated sources
follows the distribution of the {\pmnt} catalogue. For each simulated
source, we have simulated a companion source at a random angular distance
$7.04^{\prime} \leq r \leq 32.293^{\prime}$ (the FWHM values at 143 and 30\,GHz, respectively),
also following the {\pmnt} flux density
distribution. Then we have filtered the maps and have compared the filter
photometry with the input value. We find that for 
sources with flux density $S>500\,$mJy the average contamination is only 3\,\% in the worst channel case (30\,GHz). In individual cases, the contamination can be large if the source under consideration
 has an ultra-bright companion, but due to the fast decline of the number counts at high flux densities these cases are rare.
Therefore, we conclude that the contamination 
by source blending effects in the {\pmnt} is small,
and it should have little impact on the statistical
properties of the catalogue. Nevertheless, to be on the safe side, 
we provide a \texttt{Group\_FLAG} column in the catalogue,
which takes the boolean value \texttt{True}  for those sources that belong to groups, in the sense defined above. We recommend caution
when using 30-GHz flux densities around and below the 500-mJy threshold for sources with \texttt{Group\_FLAG=True}, 
especially if there is a much brighter (tens of janskys) source in the same group.

\end{appendix}

\label{lastpage}

\end{document}